\documentclass[%
 reprint,
 superscriptaddress,
preprintnumbers,
 amsmath,amssymb,
 aps,
prd,
]{revtex4-2}

\usepackage{float}
\usepackage{bbold}
\usepackage{braket}
\usepackage{amssymb,amsmath}
\usepackage{tikz}
\usepackage{bbm}
\usepackage{slashed}
\usepackage{svg}

\usepackage{amsfonts}
\usepackage{subfigure}
\usepackage{array}
\usepackage{enumerate}

\usepackage{graphicx}
\usepackage{dcolumn}
\usepackage{bm}
\usepackage[colorlinks, citecolor=blue,anchorcolor=red,menucolor=red,linkcolor=red,filecolor=red,runcolor=red,urlcolor=blue,frenchlinks=red]{hyperref}
\usepackage[nameinlink]{cleveref}
\crefformat{equation}{Eq.~(#2#1#3)}
\crefformat{figure}{Fig.~#2#1#3}
\crefformat{section}{Sec.~#2#1#3}

\usepackage{ulem}

\graphicspath{{./figures/}{./}}

\begin{document}
\preprint{RIKEN-iTHEMS-Report-25}

\title{Neural network extraction of chromo-electric and chromo-magnetic gluon masses}
\author{Jie Mei}
\email{meijie@ucas.ac.cn}
\affiliation{School of Nuclear Science and Technology, University of Chinese Academy of Sciences, Beijing, 100049,  P.R. China}
\affiliation{School of Physical Sciences, University of Chinese Academy of Sciences, Beijing 100049, P.R. China}

\author{Lingxiao Wang}
\email{lingxiao.wang@riken.jp}
\affiliation{RIKEN Interdisciplinary Theoretical and Mathematical Sciences (iTHEMS), Wako, Saitama 351-0198, Japan}

\author{Mei Huang}
\email{huangmei@ucas.ac.cn}
\affiliation{School of Nuclear Science and Technology, University of Chinese Academy of Sciences, Beijing, 100049,  P.R. China}

\begin{abstract}
    We present a neural network-based quasi-particle model to separate the contributions of chromo-electric and chromo-magnetic gluons. Using dual residual networks, we extract temperature-dependent masses from SU(3) lattice thermodynamic data of pressure and trace anomaly. After incorporating physics regularizations, the trained models reproduce lattice results with high accuracy over $T/T_c \in [1,10]$, capturing both the crossover behavior near $T_c$ and linear scaling at high temperatures. The extracted masses exhibit a physically reasonable behavior: they decrease sharply around $T_c$ and increase linearly thereafter. We find significant differences between thermal and screening masses near $T_c$, reflecting non-perturbative dynamics, while they converge at $T \gtrsim 2T_c$.
\end{abstract}
\maketitle

\section{Introduction}
Understanding the thermodynamic properties of quantum chromodynamics at finite temperature requires disentangling the contributions of different degrees of freedom in the strongly interacting medium. The quantum chromodynamics (QCD) crossover transition at $T \approx 150-170 \text{MeV}$ involves the liberation of quarks and gluons from hadronic confinement, with each type of degree of freedom contributing distinctly to the observed thermodynamic behavior~\cite{Bazavov:2014jja,Allton:2002zi}. Fermionic quark modes provide the dominant contribution to entropy and particle number density, while gluonic degrees of freedom—particularly the transverse and longitudinal polarization modes—govern the pressure and energy density through their temperature-dependent effective masses and interaction strengths~\cite{Boyd:1996bx,Peshier:1995ty,Dirks:1999uc,Meisinger:2003id}. 

Among these mentioned degrees of freedom, the distinct roles of chromo-electric and chromo-magnetic gluonic modes have been recognized as crucial for understanding the nature of the QCD phase transition. Effective theories such as the Nambu-Jona-Lasinio (NJL) model and quark-meson model, which originally incorporate only quark degrees of freedom~\cite{Hatsuda:1994pi,Klevansky:1992qe,Mei:2020jzn,Mei:2022dkd,Mei:2024rjg}, demonstrate significantly improved agreement with lattice QCD results when the Polyakov loop—representing the chromo-electric gluonic field—is included~\cite{Mei:2020jzn,Fukushima:2003fw,Ratti:2005jh,Roessner:2006xn,Schaefer:2007pw,Bhattacharyya:2012rp,Wen:2018nkn,Wen:2019ruz}, since chromo-electric gluon possesses a more significant mass change at the phase transition temperature~\cite{Xu:2011ud}. This enhancement manifests in more accurate predictions of the critical end point (CEP) location and more consistent thermodynamic behavior across the QCD phase diagram~\cite{Mei:2020jzn,Fukushima:2003fw,Ratti:2005jh,Roessner:2006xn,Schaefer:2007pw,Bhattacharyya:2012rp,Herbst:2013ufa,Gunkel:2021oya}. Chromo-magnetic gluons, on the other hand, contribute more to the pure-gauge thermodynamic properties, due to their lower effective mass compared to chromo-electric gluons. From the calculation of hard thermal loop (HTL), the chromo-electric and chromo-magnetic gluons possess the masses $m_e\sim gT$ and $m_m\sim g^2T$, accordingly~\cite{Nakamura:2003pu,Blaizot:2003tw,Kraemmer:2003gd,Andersen:2004fp,Su:2014rma}. The mass difference between chromo-electric and chromo-magnetic gluons originates from the fact that, at finite temperature, the rotational $O(4)$ symmetry is broken down to (approximate) $O(3)$ spatial symmetry since the time direction is deduced to a finite volume with $\beta = 1/T$~\cite{Kapusta:2006pm}.

To systematically account for these temperature-dependent mass differences and the resulting thermodynamic contributions from different gluonic modes, the quasiparticle model provides a particularly effective framework. This approach treats elementary particles as weakly interactive thermally modified particles with temperature-dependent effective masses, allowing for a description of the distinct behavior of chromo-electric versus chromo-magnetic degrees of freedom~\cite{Peshier:1995ty,Schneider:2001nf,Peshier:2002ww,Bluhm:2009wd,Alba:2014lda,Koothottil:2018akg}. In~\cite{Peshier:1995ty}, the author treats pure gauge system as a model of an ideal gas of quasiparticle with effective thermal mass, and use the thermodynamic properties to constrain the quasi-gluon mass. Other works also incorporate with temperature-dependent bag constant, as in~\cite{Schneider:2001nf,Peshier:2002ww,Bluhm:2009wd}. The limitation for this method lies in the fact that in traditional method, it is difficult to determine two or more mass functions with equation of states unbiasedly.

Recent advances in deep learning offer promising solutions to overcome the representation limitations. Neural networks, with their universal approximation capabilities~\cite{cybenko1989approximation,hornik1989multilayer}, can effectively handle high-dimensional parameter spaces and complex nonlinear relationships between multiple particle masses and thermodynamic observables~\cite{raissi2019physics,DL_nuclear}. The hierarchical architectures inherent in deep neural networks, combined with their automated feature extraction mechanisms, have proven particularly effective in modeling complex physical systems across diverse domains including nuclear physics~\cite{DL_nuclear,DL_nuclear1,DL_nuclear2,DL_nuclear3,DL_nuclear4,DL_nuclear5,DL_nuclear6,DL_nuclear7,DL_nuclear8,Zhou:2023pti,Chen:2025kqb,Dai:2025dir,Mansouri:2024uwc}, particle physics~\cite{DL_particle1,DL_particle2,DL_particle3,DL_particle4,DL_particle5,Wang:2023exq}, and condensed matter physics~\cite{DL_condensed1,DL_condensed2}. Recently, one new paradigm, referred to as \textit{physics-driven learning}, has been applied to the study of inverse problems in QCD physics~\cite{DL_nuclear}. As one of the applications, in the integration of machine learning with the quasiparticle model, Li et al.\cite{Li:2022ozl,Li:2025csc} have demonstrated the utility of deep neural networks in extracting temperature-dependent quasi-particle masses from the QCD equation of state. However, their formulation treats gluons as a single species, without distinguishing between chromo-electric and chromo-magnetic components. In this study, we explicitly models these two degrees of freedom and investigates their individual mass behavior. To this end, we focus on pure gauge lattice data and implement a modified quasiparticle framework that accommodates this separation. Residual networks are employed to provide sufficient representational capacity for capturing the nontrivial temperature dependence, while ensuring stable and efficient training~\cite{he2016deep,he2016identity}.

In this work, we aim to develop a systematic framework for extracting temperature-dependent effective masses of different gluonic degrees of freedom from lattice QCD data while maintaining thermodynamic consistency. To achieve this goal, we construct two individual neural networks to parameterize the mass functions of chromo-electric and chromo-magnetic gluons, $m_e(T,\theta)$ and $m_m(T,\theta)$, respectively. These neural network-derived mass functions are then incorporated into the quasiparticle model to predict the equation of state. By comparing the predicted thermodynamic quantities with lattice QCD results as in~\cite{Borsanyi:2012ve}, we construct a loss function that enables parameter optimization through backpropagation, thereby ensuring that the extracted mass functions are both physically meaningful and consistent with first-principles calculations.

This work arranges as follows: In Section~\ref{sec: method}, we present the quasiparticle model framework and dual residual neural network architecture to parameterize gluonic mass functions, along with the loss function construction including regularization terms. Section~\ref{sec: Results} demonstrates our numerical results, showing an accurate reproduction of the lattice QCD data and the extraction of physically reasonable thermal masses. Finally, Section~\ref{sec: summary} summarizes our findings and discusses potential extensions.

\section{Neural Network Extraction}\label{sec: method}

\subsection{Thermodynamics of SU(3) Gauge Theory}
For the ideal gas of non-interacting quasi-gluons, the partition function $Z$ is given by
\begin{align}
    \ln Z(T)=\ln Z_e(T)+\ln Z_m(T),
\end{align}
where the two terms in the right hand side are partition functions for chromo-electric and chromo-magnetic gluon, respectively. These two gluons follow the Bose-Einstein distribution, which gives 
\begin{align}
    &\ln Z_i(T)\nonumber\\
    &=-\frac{d_{g_i} V}{2\pi^2}\int_0^{\infty} dp\ p^2 \ln \left[1-\exp\left(-\frac{1}{T}\sqrt{p^2+m_{g_i}^2(T)}\right)\right],\label{eq:partition}
\end{align}
with $V$ being the volume of the system. $m_{g_i}(T)$ is the mass function for certain type of gluon as a function of temperature $T$. $d_{g_i}$ is the degeneracy for gluons, which is $d_{g_e}=1\times(3^2-1)=8$ for static chromo-electric gluon and $d_{g_m}=2\times(3^2-1)=16$ for chromo-magnetic gluon with 2 polarizations. 

With given partition function \cref{eq:partition}, one can get the thermodynamic properties of the system by the following relationship,
\begin{widetext}
\begin{align}
    P(T)&=T\left(\frac{\partial \ln Z(T)}{\partial V}\right)_T\nonumber\\
    &=-\sum_{i=e,m}\frac{d_{g_i} T}{2\pi^2}\int_0^{\infty} dp\ p^2 \ln \left[1-\exp\left(-\frac{1}{T}\sqrt{p^2+m_{g_i}^2(T)}\right)\right],\label{eq:qpintergal1}\\
    \varepsilon(T)&=\frac{T^2}{V}\left(\frac{\partial \ln Z(T)}{\partial T}\right)_V\nonumber\\
    &=\sum_{i=e,m}\frac{d_{g_i} T}{2\pi^2}\int_0^{\infty} dp\ p^2 \left\{\frac{\sqrt{p^2+m_{g_i}^2(T)}}{\exp\left(\sqrt{p^2+m_{g_i}^2(T)}/T\right)-1}-\frac{T}{\sqrt{p^2+m_{g_i}^2(T)}}\frac{m_{g_i}(T) \left(\partial m_{g_i}(T)/\partial T\right)}{\exp\left(\sqrt{p^2+m_{g_i}^2(T)}/T\right)-1}\right\}.
    \label{eq:qpintergal2}
\end{align}
\end{widetext}

\subsection{ResNet Quasiparticle Model}

In general, the mass functions of gluons, $m_{g_i}(T)$, can be represented as deep neural networks (DNN) with a set of trainable parameters, $\{\theta\}$. The differences between outputs of the quasiparticle models and lattice computed thermodynamical observables are used to optimize the parameters.

In particular, the momentum integral in~\cref{eq:qpintergal1,eq:qpintergal2} from $0$ to $infinity$ is computed using the 25-point \textit{Gauss-Laguerre} quadrature method to efficiently handle the exponential decay of the Bose-Einstein distribution. Once we have the temperature dependence of the mass function from DNN, the pressure and energy density above can be computed simultaneously, and we can also get the trace anomaly $\Delta(T)=(\varepsilon-3P)/T^4$.

The flowchart of the general process is shown in \cref{fig:flowchart}. We designed a Residual Neural Network (ResNet) to construct the temperature dependence of gluon masses. The network consists of two independent subnetworks, one for predicting the magnetic gluon mass ($m_m$, transverse) and the other for the electric gluon mass ($m_e$, longitudinal). Each sub-network has the following structure,

\begin{itemize}
    \item \textbf{Input Layer}: Maps the single input (temperature $T$) to a 32-dimensional hidden layer.
    \item \textbf{Residual Blocks}: Contains 7 residual blocks, each consisting of two linear layers with a \textit{Swish-like} activation function in between, defined as $f(x) = x \cdot \sigma(x)$, where $\sigma(x)$ is the sigmoid function. The residual connection is achieved by adding the input directly to the block's output.
    \item \textbf{End Layer}: Further processes the output of the residual blocks, maintaining a hidden dimension of 32.
    \item \textbf{Output Layer}: Generates non-negative mass predictions using the \textit{Softplus} activation function ($f(x) = \ln(1 + e^x)$), which guarantees the positive definiteness of gluonic masses.
\end{itemize}

From a \textit{physics-driven learning} perspective~\cite{DL_nuclear}, the \textit{Swish-like} activation function introduces nonlinearity while maintaining gradient differentiability, which helps capture complex temperature dependencies. The \textit{Softplus} activation ensures that the mass predictions are physically reasonable (non-negative).

\begin{widetext}
\centering
\begin{figure*}[htbp!]
    \centering
    \includegraphics[width=0.8\linewidth]{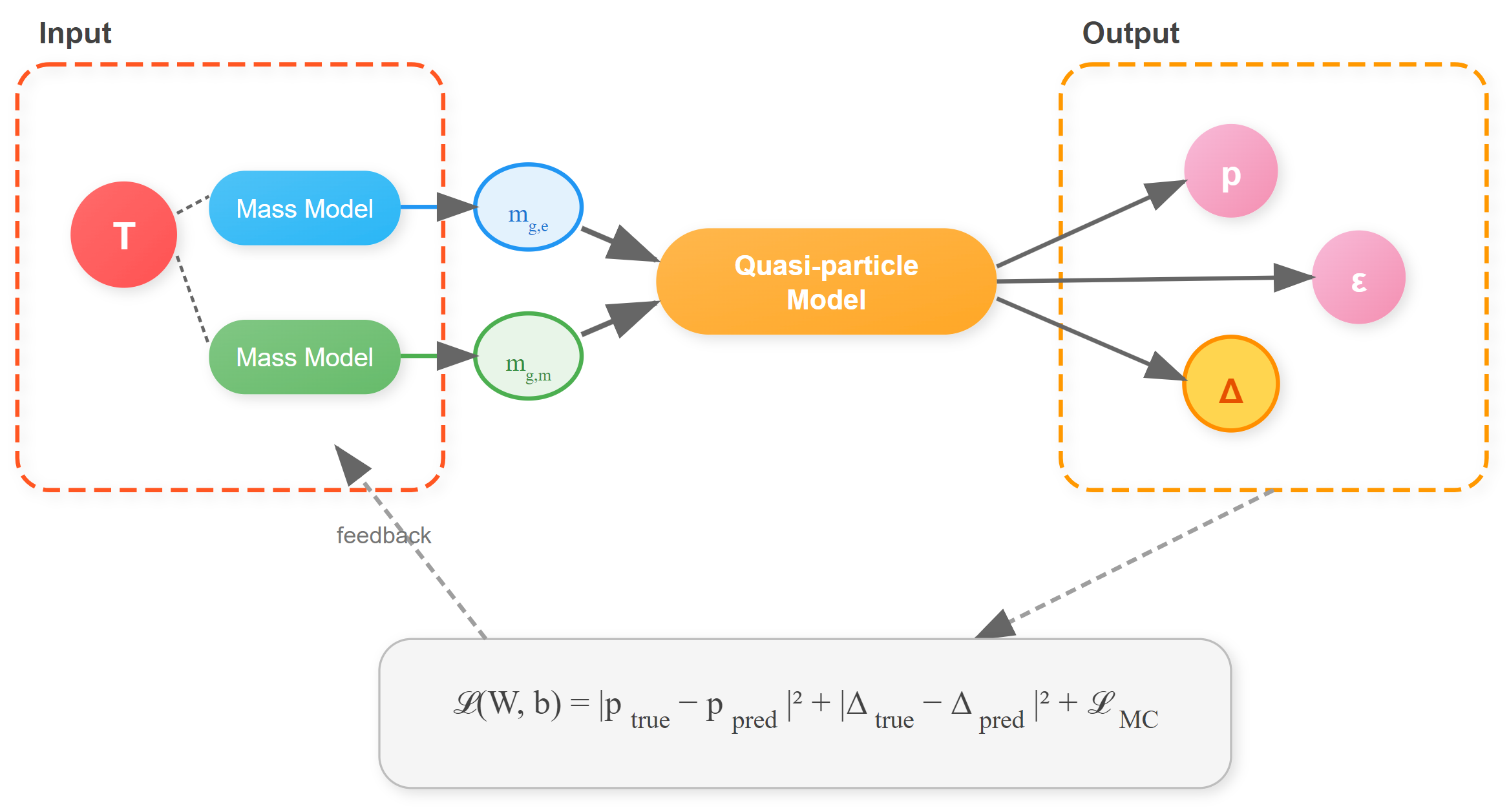}
    \caption{The flowchart of utilizing neural network to obtain the two gluonic masses is listed here. In the mass model we apply residual neural network (ResNet) and each ResNet consists of 7 hidden layers with 32 neurons per layer. Inside the residual block we utilize swish-like function as activation function.}
    \label{fig:flowchart}
\end{figure*}
\end{widetext}

With the mass functions $m_e(T, \theta)$, $m_m(T, \theta)$, we can introduce them into the quasi-particle model and calculate the corresponding thermodynamic properties. In this work, we train the neural networks mentioned before to fit the pressure and trace anomaly from lattice data. 

For the objective function, we define a custom loss function including data and regularization components,
\begin{align}
    \mathcal{L}=\mathcal{L}_{\mathrm{MSE}}+\mathcal{L}_{\mathrm{MC}}.
\end{align}

It includes, 1) \textbf{Mean Squared Error (MSE)} term, which is defined by the comparison of the predicted normalized pressure $p/T^3$ and trace anomaly $\Delta$ with the lattice data,
\begin{align}
    \mathcal{L}_{\mathrm{MSE}}=\mathrm{MSE}(p_{pred}/T^3, p_{true}/T^3)+\mathrm{MSE}(\Delta_{pred},\Delta_{true}).
\end{align}

In principle, the chi-square fitting contains statistical uncertainties, it is more appropriate to use a weighted mean squared error, where the inverse of the covariance matrix of the lattice data is incorporated as a weight. The more rigorous form of the loss term reads,
\begin{align}
\mathcal{L}_{cov} = \left(\mathbf{y}_{pred} - \mathbf{y}_{true}\right)^\top \mathbf{C}^{-1} \left(\mathbf{y}_{pred} - \mathbf{y}_{true}\right),
\end{align}
where $\mathbf{y}$ represents the observable (e.g., $p/T^3$ or $\Delta$), and $\mathbf{C}$ is the covariance matrix estimated from the lattice simulation.

However, in our case, the relative errors in the lattice data are negligibly small. Therefore, using the simplified, unweighted MSE form still leads to accurate and stable results, while reducing computational complexity. Furthermore, it is important to note that in the lattice calculations we refer to~\cite{Borsanyi:2012ve}, the pressure is not directly computed but is obtained by integrating the trace anomaly over temperature. As a result, the pressure and trace anomaly are not statistically independent. In principle, one could use only the trace anomaly as the input observable, and the model would still yield results qualitatively similar to those presented in this work. The main difference would be an increase in the uncertainty of the reconstructed mass functions, due to the loss of constraint from the direct pressure data.

2) \textbf{Regularization} term, which incorporates the expected high-temperature behavior of gluonic masses. From calculations of 3 dimensional(3D) effective theory and $N=4$ Super \textit{Yang-Mills} theory, the mass ratio $m_e/m_m$ is estimated to be around $1.8-2.2$ in the high-temperature limit. To incorporate this theoretical expectation, we introduce a soft constraint that gradually enforces the ratio $m_e/m_m \approx 2$ at sufficiently high temperatures $\mathcal{L}_{\mathrm{MC}}$. Its specific form will be given in \cref{sec:soft}, which guarantees that the model remains flexible in the low-temperature region while respecting theoretical expectations in the high temperature limit.

We have verified that the results are robust against reasonable variations in these parameter choices, confirming that our conclusions are not sensitive to the specific values selected. In the training process, we apply the \textit{Adam} optimizer with an initial learning rate of $0.0001$ which will be dynamically adjusted as the model converges. The trainable parameter $\theta$ is located on the nodes of neural network and will be updated in each epoch base on the gradient of the loss and \textit{Back-Propagation} algorithm. The training is conducted on the entire dataset (without batching) for 20,000 epochs. To assess the model’s robustness, we performed 20 independent training runs.

\section{Numerical Results}\label{sec: Results}
\begin{figure}
    \centering
    \includegraphics[width=\linewidth]{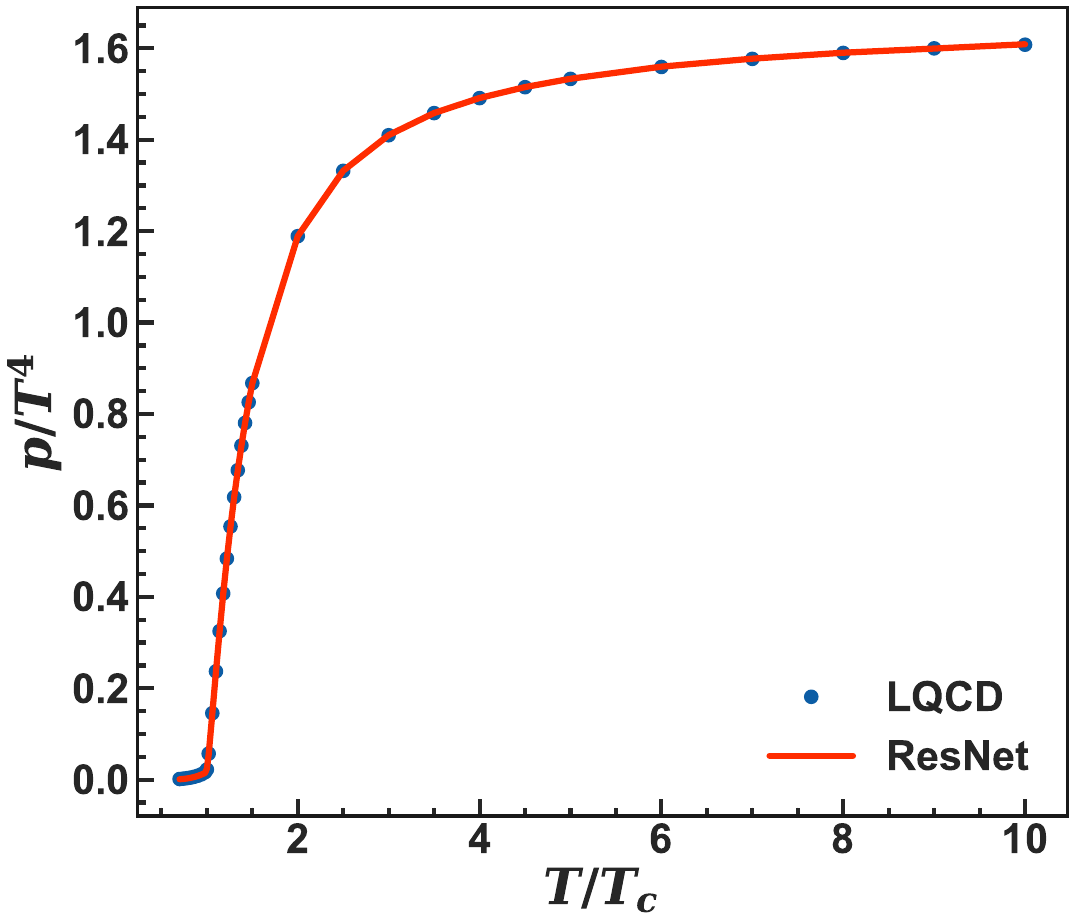}
    \includegraphics[width=\linewidth]{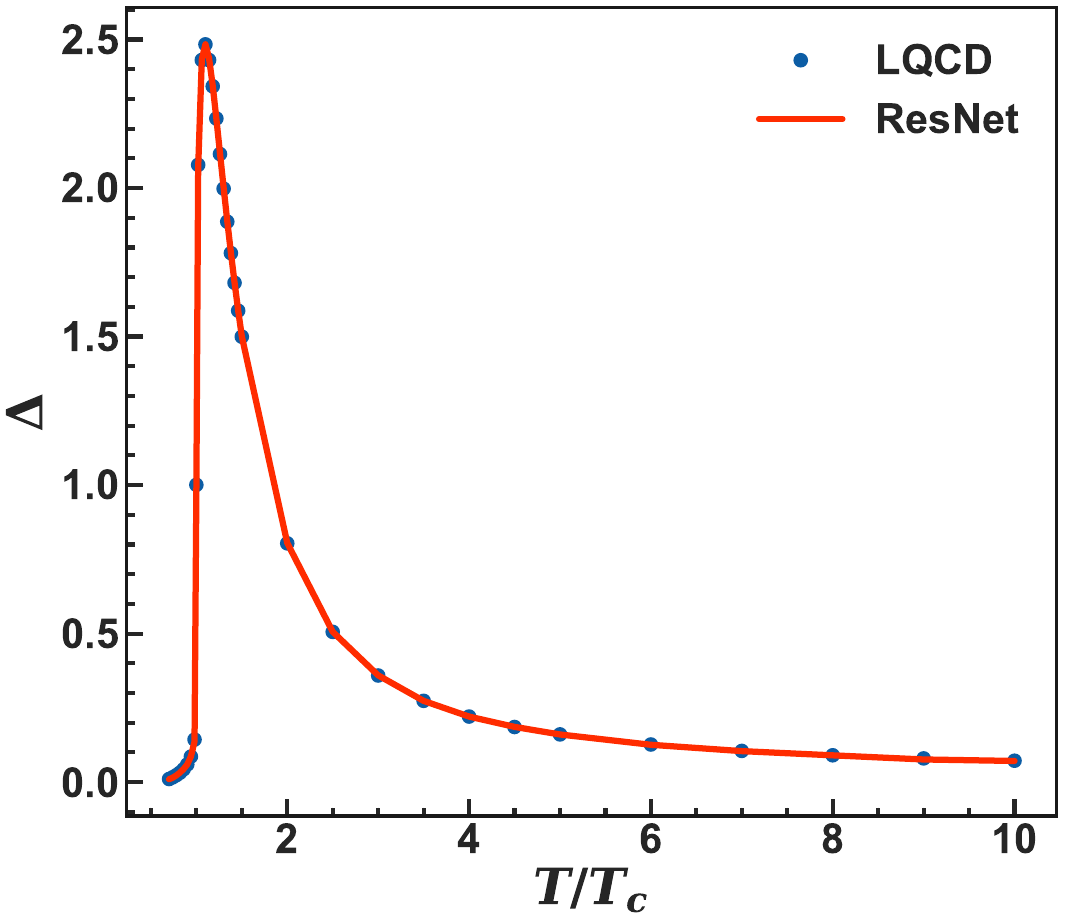}
    \caption{The results from Residual Neural Network simulation and Lattice QCD~\cite{Borsanyi:2012ve}, with critical temperature in pure-gauge theory $T_c = 0.27 \mathrm{GeV}$. The figures display the results for pressure $p/T^4$ (upper panel) and trace anomaly $\Delta$ (lower panel).}
    \label{fig:fitting}
\end{figure}
To validate the predictive accuracy of the neural network model, we compare trained output to first-principles lattice QCD data for thermodynamic observables. As shown in Figure~\ref{fig:fitting}, the trained ResNet reproduces both the pressure $p/T^4 $ and the trace anomaly $\Delta = (e - 3p)/T^4 $ with excellent precision in the whole temperature range $T/T_c \in [1, 10] $. Notably, the model accurately captures the rapid rise of $s/T^3 $ near the crossover region, as well as the sharp peak and subsequent decay of the trace anomaly, both of which reflect highly non-perturbative QCD dynamics.
\begin{figure}
    \centering
    \includegraphics[width=\linewidth]{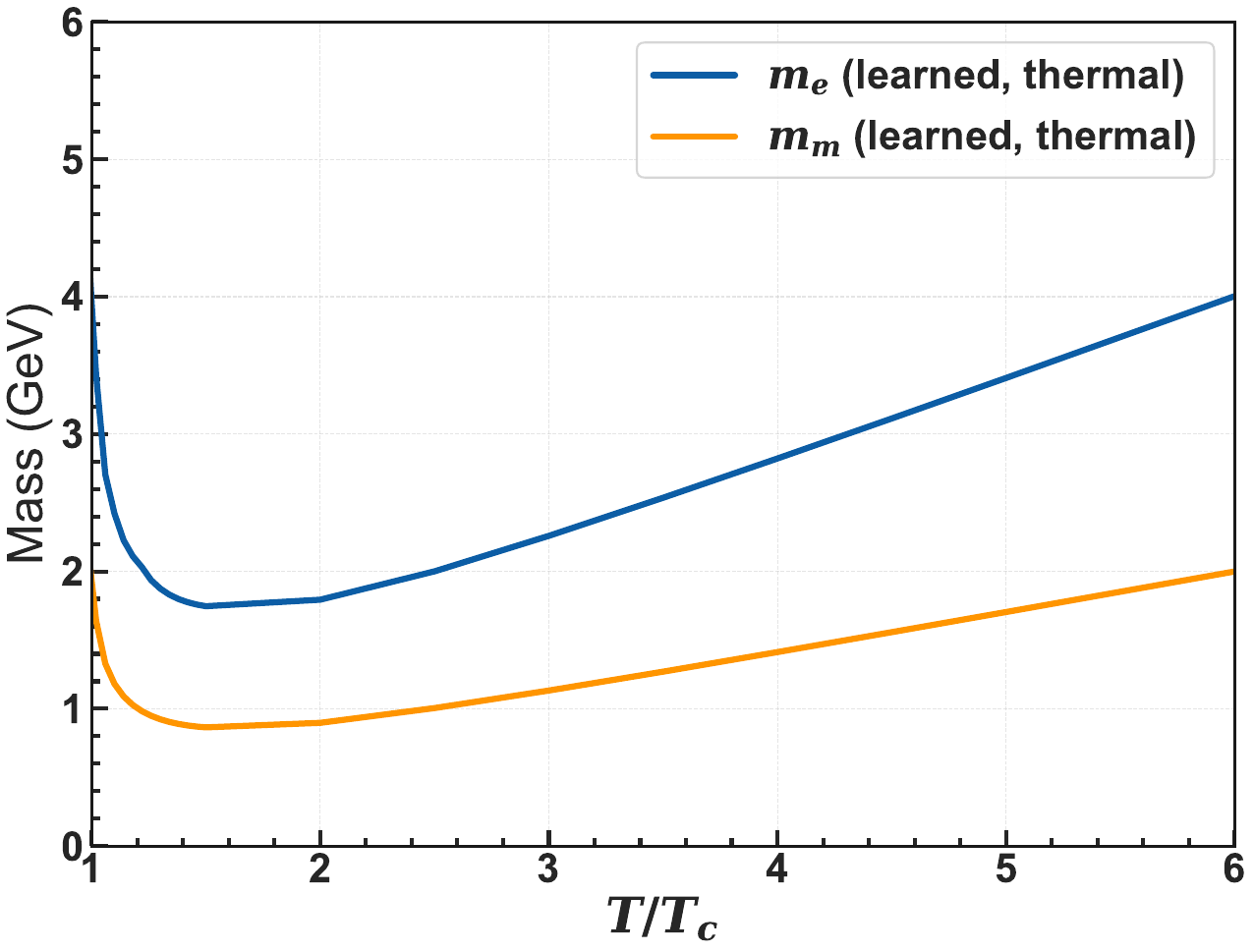}
    \caption{Chromo-electric (blue) and chromo-magnetic (orange) gluonic thermal masses, as a function of temperature, from neural network. Here we perform 20 independent runs to get the mass bands for each type of gluons.}
    \label{fig:EvM}
\end{figure}
\cref{fig:EvM} depicts the trained mass for chromo-electric and chromo-magnetic gluons separately. In the shown range, the thermal masses for two gluons possess similar behavior, where it rapidly decreases at $T\sim T_c$, reaches to its minimum, and then starts to increase linearly. In the figure we perform 20 independent runs to get the bands for the gluons. It can be seen that, once we add a proper soft regularization in the loss function, the trained gluonic masses converges well with negligible uncertainties. In the calculation of hard thermal loop, electric (longitudinal) masses stem from Debye screening of colour-electric fields while magnetic (transverse) masses reflect non-perturbative magnetic screening, suppressed by one extra power of $g$. In the high temperature region, electric and magnetic masses show a similarly linear behavior with different slope, consistent with the spirit of hard thermal loop calculation.
\begin{figure}
    \centering
    \includegraphics[width=\linewidth]{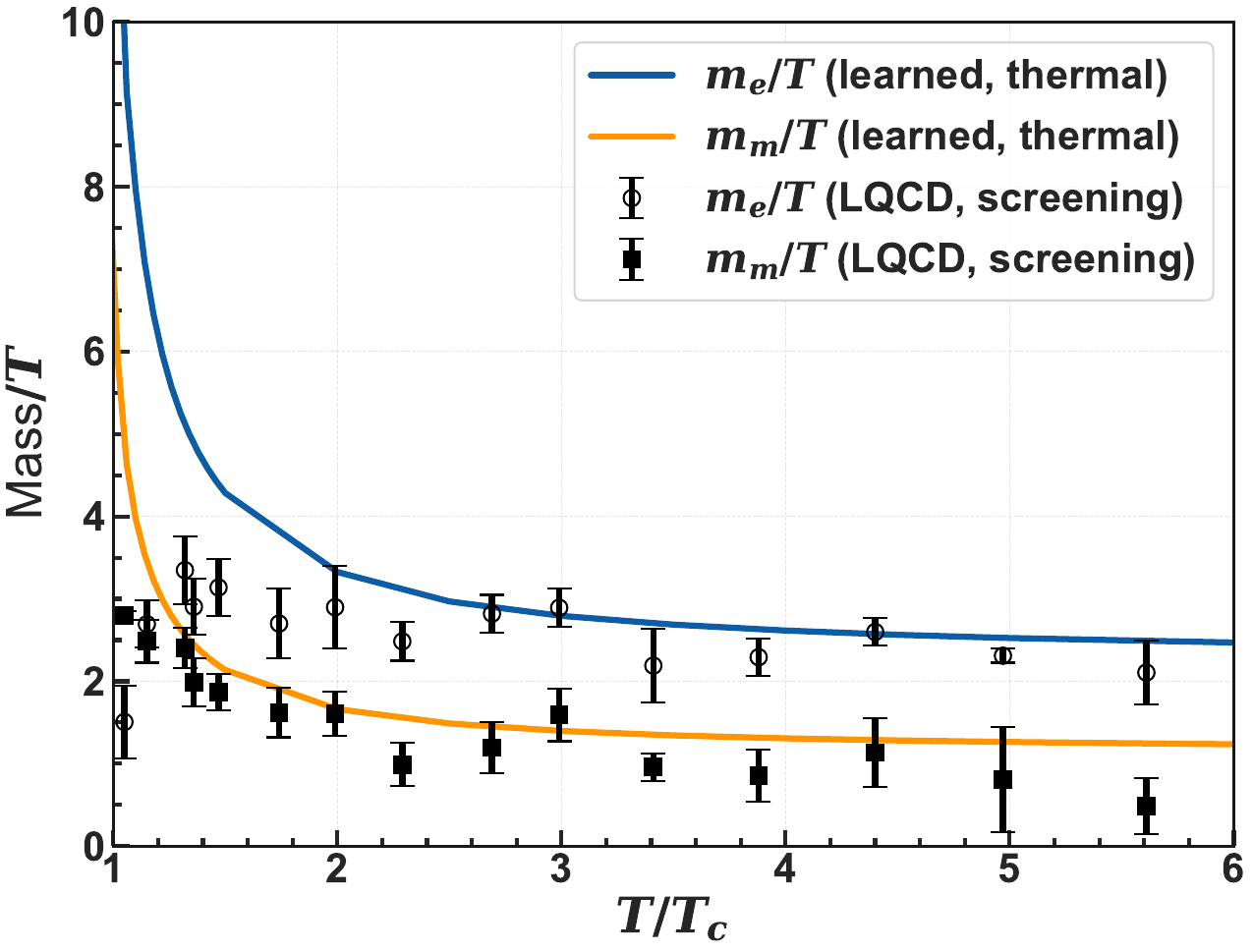}
    \caption{Chromo-electric (blue) and chromo-magnetic (orange) gluonic thermal masses divided by temperature, as a function of temperature, from neural network. Here we perform 20 independent runs to get the mass bands for each type of gluons. The lattice result for the screening masses of the corresponding gluons are also shown for comparison \cite{Nakamura:2003pu}.}
    \label{fig:EvM_over_T}
\end{figure}
\cref{fig:EvM_over_T} shows the thermal mass-over-temperature for the two types of gluons, as a function of temperature. For comparison we also display the lattice data for the screening masses for each gluons. It can be noticed that at $T\sim T_c$, the predicted gluonic masses show a huge difference to the screening masses from lattice data. This large difference between thermal and screening masses results from non-perturbative effect. As discussed in \cite{Peshier:1995ty}, thermal mass appears in the pole of a propagator, while screening mass corresponds to exponential decay of correlation function in spatial direction. At temperatures $T \gtrsim 2T_c$, the trained thermal masses become numerically close to the lattice screening masses. This apparent agreement should be interpreted with care: in the true perturbative limit one expects the chromo-electric thermal mass to be smaller than the screening mass by a factor of $\sqrt{2}$. Since $T \lesssim 6T_c$ is still far from the asymptotic scale, the observed near-coincidence mainly reflects that both quantities grow approximately linearly with $T$, rather than a genuine convergence of thermal and screening definitions.
\begin{figure}
    \centering
    \includegraphics[width=\linewidth]{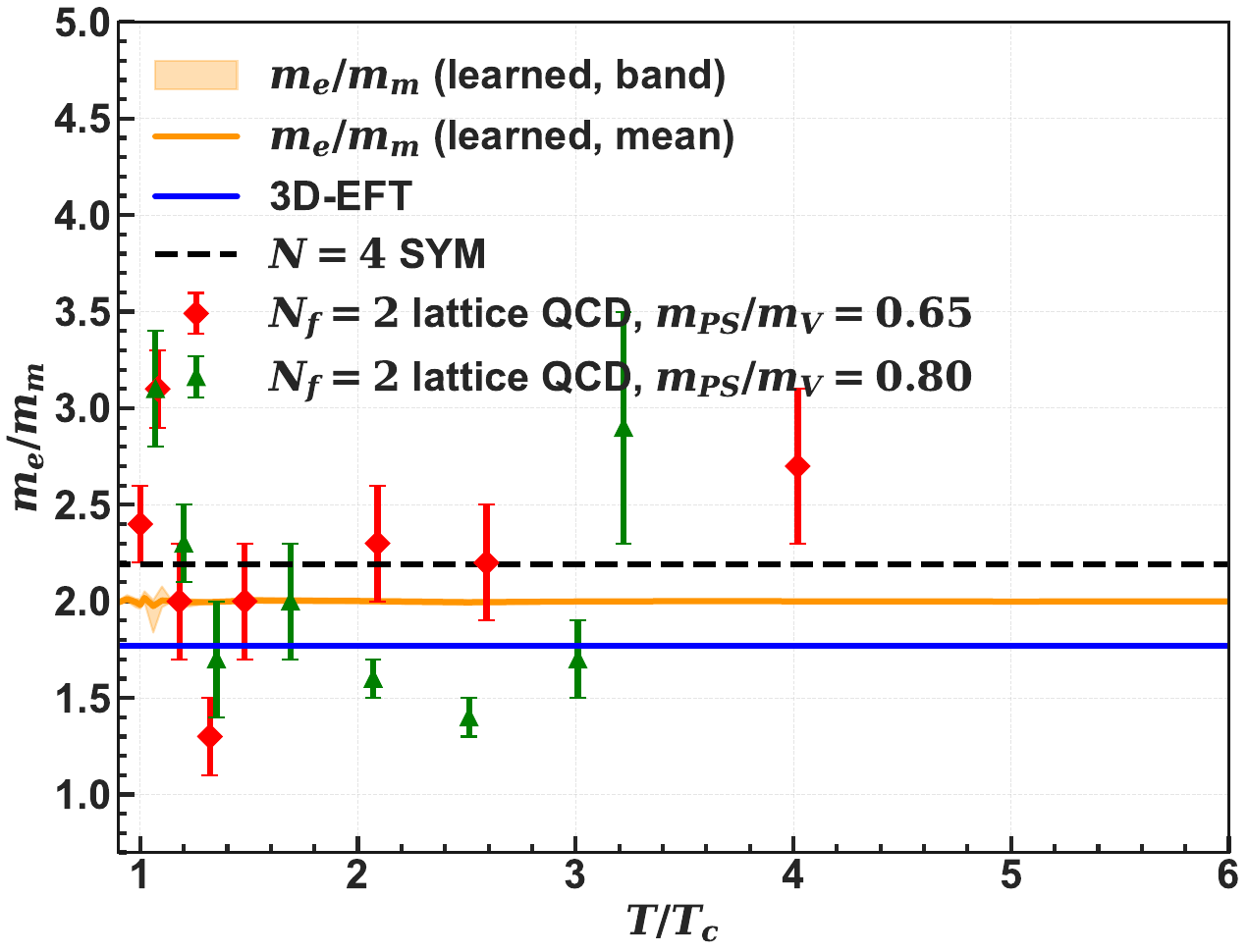}
    \caption{Ratio of chromo-electric and chromo-magnetic masses as a function of temperature. The orange band and line represent the learned prediction with uncertainty (band) and mean value (line) from a machine learning model. The blue line corresponds to the 3D effective field theory prediction~\cite{Hart:2000ha}. The dashed black line indicates the result from $\mathcal{N}=4$ Super \textit{Yang-Mills }theory~\cite{Bak:2007fk}. Lattice QCD data for two different mass ratios, $m_{PS}/m_V=0.65$ (red diamonds) and $m_{PS}/m_V=0.80$ (green triangles)~\cite{Maezawa:2010vj}, are shown with error bars.}
    \label{fig:ratio}
\end{figure}
To further assess the physical consistency of the learned quasi-particle model, we examine the ratio of the electric to magnetic screening masses $m_e/m_m $ across the deconfined temperature range, shown in Figure~\ref{fig:ratio}. The orange curve and uncertainty band represent the model prediction, trained with a soft regularization term that enforces the known asymptotic behavior $m_e/m_m \to 2 $ at high temperatures. 

Although this asymptotic constraint is built into the training objective, the resulting curve exhibits excellent compatibility with both lattice QCD results at intermediate temperatures and theoretical predictions from 3D effective field theory and $\mathcal{N} = 4 $ Super \textit{Yang-Mills} theory. This agreement reinforces the robustness of the model's interpolation and suggests that it smoothly connects non-perturbative and perturbative regimes within a unified framework.

\section{Summary and Outlook}\label{sec: summary}
In this work, we have applied deep neural networks to model the mass functions of chromo-electric and chromo-magnetic gluons separately within the quasi-particle framework. By using a dual residual network architecture with physics regularization, we were able to reproduce lattice QCD thermodynamic data with reasonable accuracies.

Our approach employs two independent neural networks to predict the temperature-dependent mass functions $m_e(T)$ and $m_m(T)$. The incorporation of a soft regularization term enforces the theoretical constraint $m_e/m_m \rightarrow 2$ at high temperatures, which is supported by perturbative calculations. The trained networks successfully reproduce both the pressure and trace anomaly from lattice QCD simulations over the temperature range $T/T_c \in [1, 10]$.

The predicted gluonic masses show reasonable physical behavior: they decrease rapidly around the crossover temperature $T_c$, reach minimum values, and then increase approximately linearly at higher temperatures. This pattern is consistent with the expected transition from non-perturbative to perturbative regimes. The mass ratio $m_e/m_m$ agrees well with available lattice data and theoretical estimates from effective field theory calculations.

An important finding is the clear distinction between the thermal masses predicted by quasi-particle model and the screening masses from lattice calculations, particularly near $T_c$. This difference reflects the non-perturbative nature of QCD in this region. At higher temperatures, the thermal and screening masses show similar magnitudes, but this is largely a pre-asymptotic feature. In the genuine perturbative regime one expects a systematic offset (e.g. $m_e^{\text{thermal}} \approx m_e^{\text{screening}}/\sqrt{2}$). Therefore, the current temperature window only suggests that the quasiparticle picture captures the right trend, rather than demonstrating full perturbative convergence.

The robustness of our results was verified through multiple independent training runs, which showed consistent convergence with small uncertainties. This suggests that the neural network approach provides a stable method for extracting gluonic mass functions from thermodynamic data. While our current study focuses on pure-gauge QCD, the methodology could be extended to include dynamical quarks or finite chemical potential effects. The approach may also be useful for investigating other aspects of QCD phenomenology where the separation of different degrees of freedom is relevant.

\section*{Acknowledgement}
We thank Drs. Tetsuo Hatsuda, Fu-Peng Li, Shuzhe Shi, Yi Yin, and Hong-An Zeng for helpful discussions.
We thank the DEEP-IN working group at RIKEN-iTHEMS for support in the preparation of this paper.
LW is supported by the RIKEN-TRIP initiative (RIKEN Quantum), JSPS KAKENHI Grant No. 25H01560, and JST-BOOST Grant; MH is supported in part by the National Natural Science Foundation of China (NSFC) Grant Nos: 12235016, 12221005.

\appendix
\section{The ``Soft'' Regularization}
\label{sec:soft}
The specific form for the regularization term is given by
\begin{align}
    \mathcal{L}_{\mathrm{MC}} = \lambda \cdot \omega(T)\left(\frac{m_e(T)}{m_m(T)}-2\right)^2
\end{align}
where $\lambda$ is a regularization strength parameter and the temperature-dependent weight $\omega(T)$ is defined as:
\begin{align}
    \omega(T) = \sigma(\alpha \cdot (T - T_{\mathrm{threshold}}))
\end{align}
with $\sigma(x)$ being the sigmoid function. The parameters are chosen based on the following physical considerations:

\begin{itemize}
\item $T_{\mathrm{threshold}} = 6T_c$ defines the temperature scale above which the high-temperature theoretical predictions become reliable
\item $\alpha = 100$ ensures a sufficiently sharp transition, preventing the constraint from significantly affecting the low-temperature regime where the theoretical ratio may not hold
\item $\lambda = 0.2$ provides a moderate regularization strength that guides the model toward the expected high-temperature behavior without overly constraining the fitting process
\end{itemize}

\begin{figure}[htbp!]
    \centering
    \includegraphics[width=\linewidth]{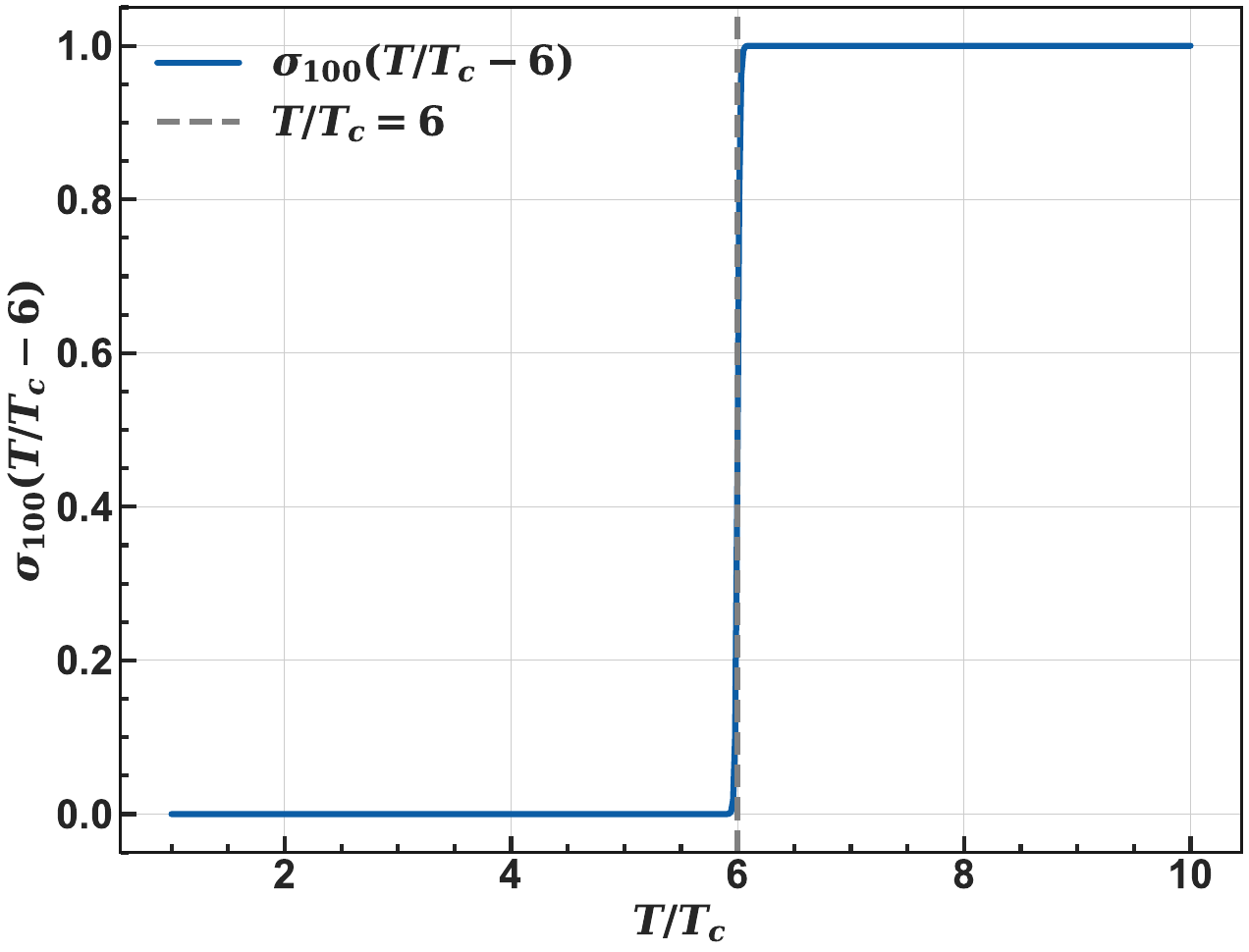}
    \caption{Temperature dependence of the weight function $\omega(T)$, showing the smooth transition from negligible constraint at low temperatures to full constraint at high temperatures.}
    \label{fig:soft_regularization}
\end{figure}

The weight function $\omega(T)$ transitions smoothly from near-zero at low temperatures to unity at high temperatures, as shown in \cref{fig:soft_regularization}. This design ensures that the model remains flexible in the low-temperature region while respecting theoretical expectations in the high-temperature limit.

\section{The Case in Absence of Regularization}

In this appendix, we present the results for the chromo-electric ($m_e$) and chromo-magnetic ($m_m$) gluon masses when the high-temperature regularization term ($\mathcal{L}_{MC}$) in the loss function is omitted. As discussed in the main text, this regularization term enforces the theoretical expectation of $m_e/m_m \approx 2$ at sufficiently high temperatures. Without this constraint, the model has greater freedom in learning the mass functions, particularly at higher temperatures.

\begin{figure}[htbp!]
    \centering
    \includegraphics[width=\columnwidth]{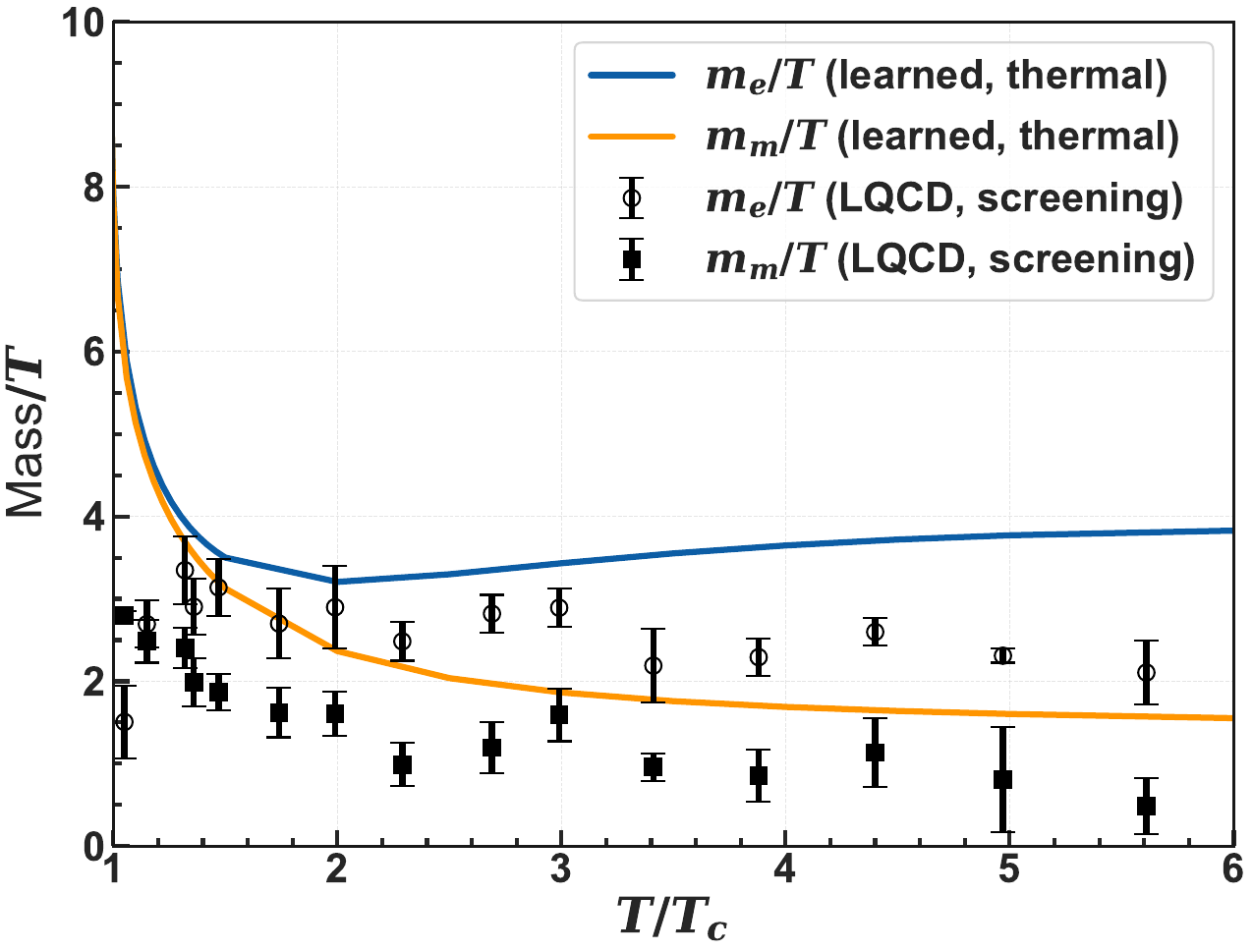}
    \caption{Chromo-electric (blue) and chromo-magnetic (orange) gluonic thermal masses, as a function of temperature, from neural network, without regularization. Only the mean from 50 independent runs is shown due to large uncertainties.}
    \label{fig:appendix_mass_no_reg}
\end{figure}

\begin{figure}[htbp!]
    \centering
    \includegraphics[width=\columnwidth]{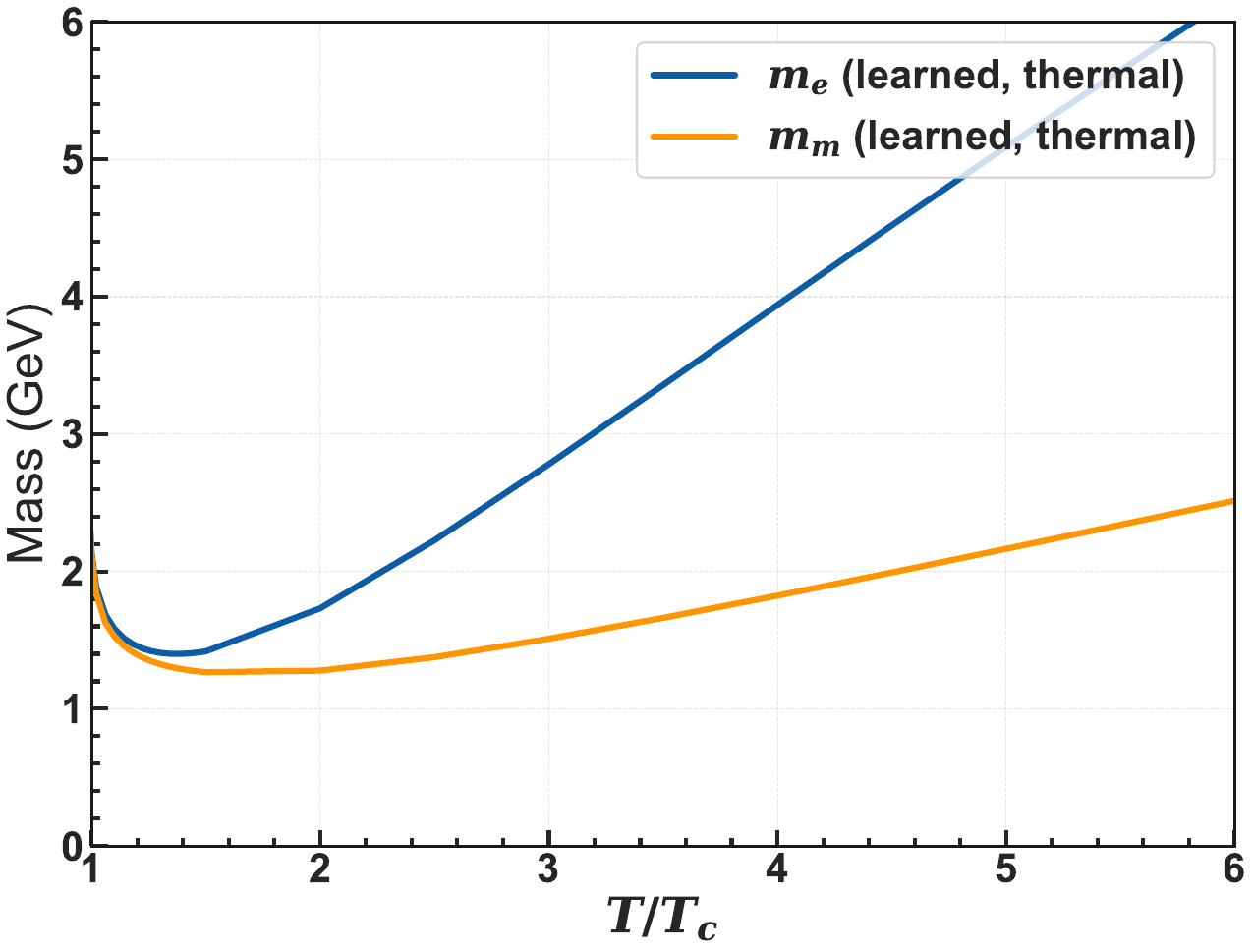}
    \caption{Chromo-electric (blue) and chromo-magnetic (orange) gluonic thermal masses divided by temperature, as a function of temperature, from neural network, without regularization. Lattice screening masses are also shown for comparison. Only the mean from 50 independent runs is shown.}
    \label{fig:appendix_mass_over_T_no_reg}
\end{figure}

Figure \ref{fig:appendix_mass_no_reg} (similar to Figure 4 in the main text, but without regularization) shows the temperature-dependent thermal masses of chromo-electric and chromo-magnetic gluons. Figure \ref{fig:appendix_mass_over_T_no_reg} (similar to Figure 5 in the main text, but without regularization) displays the ratio of these thermal masses to temperature, alongside lattice QCD screening mass data. Due to the absence of the high-temperature regularization, the trained mass bands exhibit significantly larger uncertainties compared to the results presented in the main text. For clarity, only the mean values from 50 independent training runs are shown in these figures, as the width of the mass bands becomes prohibitively large without the regularization.
Despite the increased uncertainty, these results demonstrate that the neural network approach can still distinguish between chromo-electric and chromo-magnetic gluon contributions and learn their separate temperature-dependent masses even without the explicit high-temperature regularization. However, the resulting mass functions are less constrained and appear more "disordered" at higher temperatures. This highlights the crucial role of the regularization term in guiding the model towards physically meaningful and well-behaved solutions, especially in the high-temperature regime where theoretical predictions for the mass ratio are available. The regularization effectively incorporates prior physics knowledge, leading to more robust and interpretable results for the gluon masses.

\section{Calculation of Shear Viscosity}
In addition to reproducing thermodynamic quantities, the quasiparticle model allows us to estimate transport coefficients. In this section, we illustrate the shear viscosity to entropy density ratio, $\eta/s$, for the pure gauge system, using the temperature-dependent gluonic masses extracted from the machine learning model.

The starting point is the relaxation time approximation (RTA) to the Boltzmann equation~\cite{Plumari:2011mk,Li:2022ozl}, where the shear viscosity is expressed as
\begin{equation}
\eta = \frac{1}{15T} \sum_{i=e,m} d_i \int \frac{d^3p}{(2\pi)^3} \frac{p^4}{E_i^2} \, \tau_i \, f_i(1+f_i),
\end{equation}
with $E_i = \sqrt{p^2 + m_i^2(T)}$ being the quasiparticle energy of chromo-electric ($i=e$) and chromo-magnetic ($i=m$) gluons, and $f_i = (e^{E_i/T}-1)^{-1}$ their Bose–Einstein distribution. The relaxation time $\tau_i$ encodes the microscopic interaction rate. In practice, we employ the perturbatively motivated \textit{Ansatz},
\begin{equation}
\tau_i^{-1} = \frac{N_c}{4\pi}\, g^2(T) \, T \ln\!\left(\frac{2k}{g^2(T)}\right),
\end{equation}
with $N_c=3$, and $k$ a phenomenological constant fixed at $k = 5.4$ to approach to Kovtun-Son-Starinets (KSS) bound $1/4\pi$~\cite{Kovtun:2004de} at its minimum. The running coupling $g^2(T)$ is determined self-consistently from the quasiparticle thermal masses, since in hard-thermal-loop calculations one expects $m_e^2 = \frac{N_c}{6}g^2T^2$.

The entropy density $s$ is not evaluated from the quasiparticle model itself, but instead obtained directly from lattice QCD results of pressure $p$ and trace anomaly $\Delta$ through
\begin{equation}
s(T) = \frac{\epsilon(T)+p(T)}{T} \,,
\qquad
\epsilon(T) = \Delta(T)\,T^4 + 3p(T),
\end{equation}
where $p(T)$ and $\Delta(T)$ are taken from the continuum-extrapolated SU(3) lattice data.

\begin{figure}[t]
\centering
\includegraphics[width=\columnwidth]{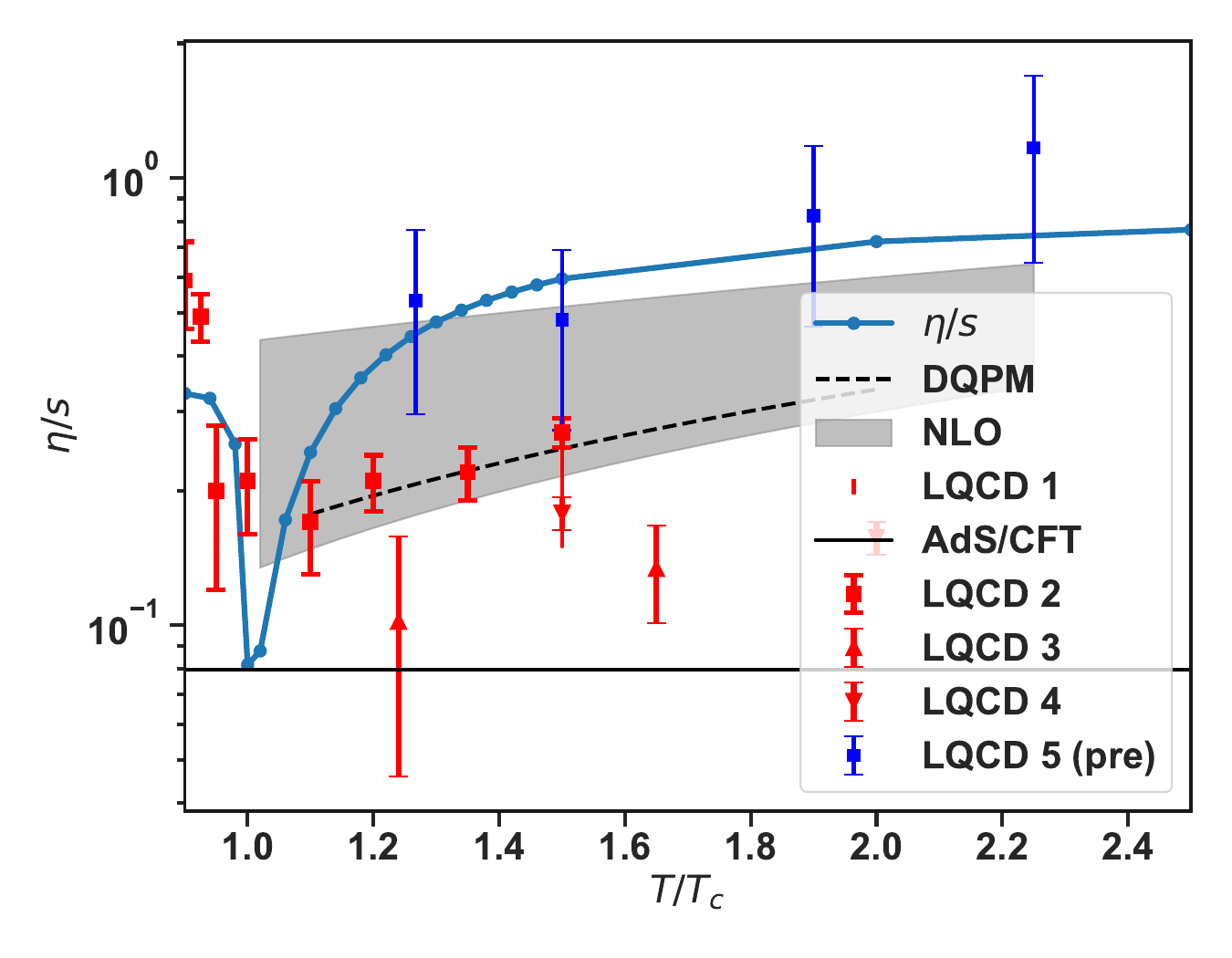}
\caption{The shear viscosity to entropy density ratio $\eta/s$ from the quasiparticle model with machine-learning extracted gluon masses (blue line), compared with various lattice QCD results (LQCD 1 ~\cite{Altenkort:2022yhb}, LQCD 2~\cite{Astrakhantsev:2017nrs}, LQCD 3~\cite{Meyer:2007ic}, LQCD 4~\cite{Borsanyi:2018srz}, LQCD 5~\cite{Zhang2025}) and perturbative estimates (DQPM~\cite{Marty:2013ita}, NLO~\cite{Ghiglieri:2018dib}).}
\label{fig:etas}
\end{figure}

Numerically, the momentum integrals are performed using \textit{Gauss–Laguerre} quadrature, which is efficient for the exponentially decaying Bose–Einstein factor. The results for $\eta/s$ are shown in Fig.~\ref{fig:etas}, together with available lattice QCD estimates~\cite{Altenkort:2022yhb,Astrakhantsev:2017nrs,Meyer:2007ic,Borsanyi:2018srz,Zhang2025} and perturbative calculations~\cite{Marty:2013ita, Ghiglieri:2018dib} for comparison. We find that $\eta/s$ exhibits a pronounced minimum close to $T_c$, reaching values close to the conjectured AdS/CFT bound of $1/4\pi$~\cite{Kovtun:2004de}, and then increases monotonically with temperature. The magnitude and trend are consistent with existing lattice data, especially from LQCD 5~\cite{Zhang2025}, within uncertainties.
\bibliography{apssamp}

\begin{thebibliography}{73}%
\makeatletter
\providecommand \@ifxundefined [1]{%
 \@ifx{#1\undefined}
}%
\providecommand \@ifnum [1]{%
 \ifnum #1\expandafter \@firstoftwo
 \else \expandafter \@secondoftwo
 \fi
}%
\providecommand \@ifx [1]{%
 \ifx #1\expandafter \@firstoftwo
 \else \expandafter \@secondoftwo
 \fi
}%
\providecommand \natexlab [1]{#1}%
\providecommand \enquote  [1]{``#1''}%
\providecommand \bibnamefont  [1]{#1}%
\providecommand \bibfnamefont [1]{#1}%
\providecommand \citenamefont [1]{#1}%
\providecommand \href@noop [0]{\@secondoftwo}%
\providecommand \href [0]{\begingroup \@sanitize@url \@href}%
\providecommand \@href[1]{\@@startlink{#1}\@@href}%
\providecommand \@@href[1]{\endgroup#1\@@endlink}%
\providecommand \@sanitize@url [0]{\catcode `\\12\catcode `\$12\catcode `\&12\catcode `\#12\catcode `\^12\catcode `\_12\catcode `\%12\relax}%
\providecommand \@@startlink[1]{}%
\providecommand \@@endlink[0]{}%
\providecommand \url  [0]{\begingroup\@sanitize@url \@url }%
\providecommand \@url [1]{\endgroup\@href {#1}{\urlprefix }}%
\providecommand \urlprefix  [0]{URL }%
\providecommand \Eprint [0]{\href }%
\providecommand \doibase [0]{https://doi.org/}%
\providecommand \selectlanguage [0]{\@gobble}%
\providecommand \bibinfo  [0]{\@secondoftwo}%
\providecommand \bibfield  [0]{\@secondoftwo}%
\providecommand \translation [1]{[#1]}%
\providecommand \BibitemOpen [0]{}%
\providecommand \bibitemStop [0]{}%
\providecommand \bibitemNoStop [0]{.\EOS\space}%
\providecommand \EOS [0]{\spacefactor3000\relax}%
\providecommand \BibitemShut  [1]{\csname bibitem#1\endcsname}%
\let\auto@bib@innerbib\@empty
\bibitem [{\citenamefont {Bazavov}(2014)}]{Bazavov:2014jja}%
  \BibitemOpen
  \bibfield  {author} {\bibinfo {author} {\bibfnamefont {A.}~\bibnamefont {Bazavov}} (\bibinfo {collaboration} {HotQCD}),\ }\bibfield  {title} {\bibinfo {title} {{The QCD equation of state}},\ }\href {https://doi.org/10.1016/j.nuclphysa.2014.08.073} {\bibfield  {journal} {\bibinfo  {journal} {Nucl. Phys. A}\ }\textbf {\bibinfo {volume} {931}},\ \bibinfo {pages} {867} (\bibinfo {year} {2014})}\BibitemShut {NoStop}%
\bibitem [{\citenamefont {Allton}\ \emph {et~al.}(2002)\citenamefont {Allton}, \citenamefont {Ejiri}, \citenamefont {Hands}, \citenamefont {Kaczmarek}, \citenamefont {Karsch}, \citenamefont {Laermann}, \citenamefont {Schmidt},\ and\ \citenamefont {Scorzato}}]{Allton:2002zi}%
  \BibitemOpen
  \bibfield  {author} {\bibinfo {author} {\bibfnamefont {C.~R.}\ \bibnamefont {Allton}}, \bibinfo {author} {\bibfnamefont {S.}~\bibnamefont {Ejiri}}, \bibinfo {author} {\bibfnamefont {S.~J.}\ \bibnamefont {Hands}}, \bibinfo {author} {\bibfnamefont {O.}~\bibnamefont {Kaczmarek}}, \bibinfo {author} {\bibfnamefont {F.}~\bibnamefont {Karsch}}, \bibinfo {author} {\bibfnamefont {E.}~\bibnamefont {Laermann}}, \bibinfo {author} {\bibfnamefont {C.}~\bibnamefont {Schmidt}},\ and\ \bibinfo {author} {\bibfnamefont {L.}~\bibnamefont {Scorzato}},\ }\bibfield  {title} {\bibinfo {title} {{The QCD thermal phase transition in the presence of a small chemical potential}},\ }\href {https://doi.org/10.1103/PhysRevD.66.074507} {\bibfield  {journal} {\bibinfo  {journal} {Phys. Rev. D}\ }\textbf {\bibinfo {volume} {66}},\ \bibinfo {pages} {074507} (\bibinfo {year} {2002})},\ \Eprint {https://arxiv.org/abs/hep-lat/0204010} {arXiv:hep-lat/0204010} \BibitemShut {NoStop}%
\bibitem [{\citenamefont {Boyd}\ \emph {et~al.}(1996)\citenamefont {Boyd}, \citenamefont {Engels}, \citenamefont {Karsch}, \citenamefont {Laermann}, \citenamefont {Legeland}, \citenamefont {Lutgemeier},\ and\ \citenamefont {Petersson}}]{Boyd:1996bx}%
  \BibitemOpen
  \bibfield  {author} {\bibinfo {author} {\bibfnamefont {G.}~\bibnamefont {Boyd}}, \bibinfo {author} {\bibfnamefont {J.}~\bibnamefont {Engels}}, \bibinfo {author} {\bibfnamefont {F.}~\bibnamefont {Karsch}}, \bibinfo {author} {\bibfnamefont {E.}~\bibnamefont {Laermann}}, \bibinfo {author} {\bibfnamefont {C.}~\bibnamefont {Legeland}}, \bibinfo {author} {\bibfnamefont {M.}~\bibnamefont {Lutgemeier}},\ and\ \bibinfo {author} {\bibfnamefont {B.}~\bibnamefont {Petersson}},\ }\bibfield  {title} {\bibinfo {title} {{Thermodynamics of SU(3) lattice gauge theory}},\ }\href {https://doi.org/10.1016/0550-3213(96)00170-8} {\bibfield  {journal} {\bibinfo  {journal} {Nucl. Phys. B}\ }\textbf {\bibinfo {volume} {469}},\ \bibinfo {pages} {419} (\bibinfo {year} {1996})},\ \Eprint {https://arxiv.org/abs/hep-lat/9602007} {arXiv:hep-lat/9602007} \BibitemShut {NoStop}%
\bibitem [{\citenamefont {Peshier}\ \emph {et~al.}(1996)\citenamefont {Peshier}, \citenamefont {Kampfer}, \citenamefont {Pavlenko},\ and\ \citenamefont {Soff}}]{Peshier:1995ty}%
  \BibitemOpen
  \bibfield  {author} {\bibinfo {author} {\bibfnamefont {A.}~\bibnamefont {Peshier}}, \bibinfo {author} {\bibfnamefont {B.}~\bibnamefont {Kampfer}}, \bibinfo {author} {\bibfnamefont {O.~P.}\ \bibnamefont {Pavlenko}},\ and\ \bibinfo {author} {\bibfnamefont {G.}~\bibnamefont {Soff}},\ }\bibfield  {title} {\bibinfo {title} {{A Massive quasiparticle model of the SU(3) gluon plasma}},\ }\href {https://doi.org/10.1103/PhysRevD.54.2399} {\bibfield  {journal} {\bibinfo  {journal} {Phys. Rev. D}\ }\textbf {\bibinfo {volume} {54}},\ \bibinfo {pages} {2399} (\bibinfo {year} {1996})}\BibitemShut {NoStop}%
\bibitem [{\citenamefont {Dirks}\ \emph {et~al.}(1999)\citenamefont {Dirks}, \citenamefont {Niegawa},\ and\ \citenamefont {Okano}}]{Dirks:1999uc}%
  \BibitemOpen
  \bibfield  {author} {\bibinfo {author} {\bibfnamefont {M.}~\bibnamefont {Dirks}}, \bibinfo {author} {\bibfnamefont {A.}~\bibnamefont {Niegawa}},\ and\ \bibinfo {author} {\bibfnamefont {K.}~\bibnamefont {Okano}},\ }\bibfield  {title} {\bibinfo {title} {{A Slavnov-Taylor identity and equality of damping rates for static transverse and longitudinal gluons in hot QCD}},\ }\href {https://doi.org/10.1016/S0370-2693(99)00829-1} {\bibfield  {journal} {\bibinfo  {journal} {Phys. Lett. B}\ }\textbf {\bibinfo {volume} {461}},\ \bibinfo {pages} {131} (\bibinfo {year} {1999})},\ \Eprint {https://arxiv.org/abs/hep-ph/9907439} {arXiv:hep-ph/9907439} \BibitemShut {NoStop}%
\bibitem [{\citenamefont {Meisinger}\ \emph {et~al.}(2004)\citenamefont {Meisinger}, \citenamefont {Ogilvie},\ and\ \citenamefont {Miller}}]{Meisinger:2003id}%
  \BibitemOpen
  \bibfield  {author} {\bibinfo {author} {\bibfnamefont {P.~N.}\ \bibnamefont {Meisinger}}, \bibinfo {author} {\bibfnamefont {M.~C.}\ \bibnamefont {Ogilvie}},\ and\ \bibinfo {author} {\bibfnamefont {T.~R.}\ \bibnamefont {Miller}},\ }\bibfield  {title} {\bibinfo {title} {{Gluon quasiparticles and the polyakov loop}},\ }\href {https://doi.org/10.1016/j.physletb.2004.02.009} {\bibfield  {journal} {\bibinfo  {journal} {Phys. Lett. B}\ }\textbf {\bibinfo {volume} {585}},\ \bibinfo {pages} {149} (\bibinfo {year} {2004})},\ \Eprint {https://arxiv.org/abs/hep-ph/0312272} {arXiv:hep-ph/0312272} \BibitemShut {NoStop}%
\bibitem [{\citenamefont {Hatsuda}\ and\ \citenamefont {Kunihiro}(1994)}]{Hatsuda:1994pi}%
  \BibitemOpen
  \bibfield  {author} {\bibinfo {author} {\bibfnamefont {T.}~\bibnamefont {Hatsuda}}\ and\ \bibinfo {author} {\bibfnamefont {T.}~\bibnamefont {Kunihiro}},\ }\bibfield  {title} {\bibinfo {title} {{QCD phenomenology based on a chiral effective Lagrangian}},\ }\href {https://doi.org/10.1016/0370-1573(94)90022-1} {\bibfield  {journal} {\bibinfo  {journal} {Phys. Rept.}\ }\textbf {\bibinfo {volume} {247}},\ \bibinfo {pages} {221} (\bibinfo {year} {1994})},\ \Eprint {https://arxiv.org/abs/hep-ph/9401310} {arXiv:hep-ph/9401310} \BibitemShut {NoStop}%
\bibitem [{\citenamefont {Klevansky}(1992)}]{Klevansky:1992qe}%
  \BibitemOpen
  \bibfield  {author} {\bibinfo {author} {\bibfnamefont {S.~P.}\ \bibnamefont {Klevansky}},\ }\bibfield  {title} {\bibinfo {title} {{The Nambu-Jona-Lasinio model of quantum chromodynamics}},\ }\href {https://doi.org/10.1103/RevModPhys.64.649} {\bibfield  {journal} {\bibinfo  {journal} {Rev. Mod. Phys.}\ }\textbf {\bibinfo {volume} {64}},\ \bibinfo {pages} {649} (\bibinfo {year} {1992})}\BibitemShut {NoStop}%
\bibitem [{\citenamefont {Mei}\ and\ \citenamefont {Mao}(2020)}]{Mei:2020jzn}%
  \BibitemOpen
  \bibfield  {author} {\bibinfo {author} {\bibfnamefont {J.}~\bibnamefont {Mei}}\ and\ \bibinfo {author} {\bibfnamefont {S.}~\bibnamefont {Mao}},\ }\bibfield  {title} {\bibinfo {title} {{Inverse catalysis effect of the quark anomalous magnetic moment to chiral restoration and deconfinement phase transitions}},\ }\href {https://doi.org/10.1103/PhysRevD.102.114035} {\bibfield  {journal} {\bibinfo  {journal} {Phys. Rev. D}\ }\textbf {\bibinfo {volume} {102}},\ \bibinfo {pages} {114035} (\bibinfo {year} {2020})},\ \Eprint {https://arxiv.org/abs/2008.12123} {arXiv:2008.12123 [hep-ph]} \BibitemShut {NoStop}%
\bibitem [{\citenamefont {Mei}\ \emph {et~al.}(2023)\citenamefont {Mei}, \citenamefont {Xia},\ and\ \citenamefont {Mao}}]{Mei:2022dkd}%
  \BibitemOpen
  \bibfield  {author} {\bibinfo {author} {\bibfnamefont {J.}~\bibnamefont {Mei}}, \bibinfo {author} {\bibfnamefont {T.}~\bibnamefont {Xia}},\ and\ \bibinfo {author} {\bibfnamefont {S.}~\bibnamefont {Mao}},\ }\bibfield  {title} {\bibinfo {title} {{Mass spectra of neutral mesons K0,\ensuremath{\pi}0,\ensuremath{\eta},\ensuremath{\eta}' at finite magnetic field, temperature and quark chemical potential}},\ }\href {https://doi.org/10.1103/PhysRevD.107.074018} {\bibfield  {journal} {\bibinfo  {journal} {Phys. Rev. D}\ }\textbf {\bibinfo {volume} {107}},\ \bibinfo {pages} {074018} (\bibinfo {year} {2023})},\ \Eprint {https://arxiv.org/abs/2212.04778} {arXiv:2212.04778 [hep-ph]} \BibitemShut {NoStop}%
\bibitem [{\citenamefont {Mei}\ \emph {et~al.}(2024)\citenamefont {Mei}, \citenamefont {Wen}, \citenamefont {Mao}, \citenamefont {Huang},\ and\ \citenamefont {Xu}}]{Mei:2024rjg}%
  \BibitemOpen
  \bibfield  {author} {\bibinfo {author} {\bibfnamefont {J.}~\bibnamefont {Mei}}, \bibinfo {author} {\bibfnamefont {R.}~\bibnamefont {Wen}}, \bibinfo {author} {\bibfnamefont {S.}~\bibnamefont {Mao}}, \bibinfo {author} {\bibfnamefont {M.}~\bibnamefont {Huang}},\ and\ \bibinfo {author} {\bibfnamefont {K.}~\bibnamefont {Xu}},\ }\bibfield  {title} {\bibinfo {title} {{Magnetic catalysis and diamagnetism from pion fluctuations}},\ }\href {https://doi.org/10.1103/PhysRevD.110.034024} {\bibfield  {journal} {\bibinfo  {journal} {Phys. Rev. D}\ }\textbf {\bibinfo {volume} {110}},\ \bibinfo {pages} {034024} (\bibinfo {year} {2024})},\ \Eprint {https://arxiv.org/abs/2402.19193} {arXiv:2402.19193 [hep-ph]} \BibitemShut {NoStop}%
\bibitem [{\citenamefont {Fukushima}(2004)}]{Fukushima:2003fw}%
  \BibitemOpen
  \bibfield  {author} {\bibinfo {author} {\bibfnamefont {K.}~\bibnamefont {Fukushima}},\ }\bibfield  {title} {\bibinfo {title} {{Chiral effective model with the Polyakov loop}},\ }\href {https://doi.org/10.1016/j.physletb.2004.04.027} {\bibfield  {journal} {\bibinfo  {journal} {Phys. Lett. B}\ }\textbf {\bibinfo {volume} {591}},\ \bibinfo {pages} {277} (\bibinfo {year} {2004})},\ \Eprint {https://arxiv.org/abs/hep-ph/0310121} {arXiv:hep-ph/0310121} \BibitemShut {NoStop}%
\bibitem [{\citenamefont {Ratti}\ \emph {et~al.}(2006)\citenamefont {Ratti}, \citenamefont {Thaler},\ and\ \citenamefont {Weise}}]{Ratti:2005jh}%
  \BibitemOpen
  \bibfield  {author} {\bibinfo {author} {\bibfnamefont {C.}~\bibnamefont {Ratti}}, \bibinfo {author} {\bibfnamefont {M.~A.}\ \bibnamefont {Thaler}},\ and\ \bibinfo {author} {\bibfnamefont {W.}~\bibnamefont {Weise}},\ }\bibfield  {title} {\bibinfo {title} {{Phases of QCD: Lattice thermodynamics and a field theoretical model}},\ }\href {https://doi.org/10.1103/PhysRevD.73.014019} {\bibfield  {journal} {\bibinfo  {journal} {Phys. Rev. D}\ }\textbf {\bibinfo {volume} {73}},\ \bibinfo {pages} {014019} (\bibinfo {year} {2006})},\ \Eprint {https://arxiv.org/abs/hep-ph/0506234} {arXiv:hep-ph/0506234} \BibitemShut {NoStop}%
\bibitem [{\citenamefont {Roessner}\ \emph {et~al.}(2007)\citenamefont {Roessner}, \citenamefont {Ratti},\ and\ \citenamefont {Weise}}]{Roessner:2006xn}%
  \BibitemOpen
  \bibfield  {author} {\bibinfo {author} {\bibfnamefont {S.}~\bibnamefont {Roessner}}, \bibinfo {author} {\bibfnamefont {C.}~\bibnamefont {Ratti}},\ and\ \bibinfo {author} {\bibfnamefont {W.}~\bibnamefont {Weise}},\ }\bibfield  {title} {\bibinfo {title} {{Polyakov loop, diquarks and the two-flavour phase diagram}},\ }\href {https://doi.org/10.1103/PhysRevD.75.034007} {\bibfield  {journal} {\bibinfo  {journal} {Phys. Rev. D}\ }\textbf {\bibinfo {volume} {75}},\ \bibinfo {pages} {034007} (\bibinfo {year} {2007})},\ \Eprint {https://arxiv.org/abs/hep-ph/0609281} {arXiv:hep-ph/0609281} \BibitemShut {NoStop}%
\bibitem [{\citenamefont {Schaefer}\ \emph {et~al.}(2007)\citenamefont {Schaefer}, \citenamefont {Pawlowski},\ and\ \citenamefont {Wambach}}]{Schaefer:2007pw}%
  \BibitemOpen
  \bibfield  {author} {\bibinfo {author} {\bibfnamefont {B.-J.}\ \bibnamefont {Schaefer}}, \bibinfo {author} {\bibfnamefont {J.~M.}\ \bibnamefont {Pawlowski}},\ and\ \bibinfo {author} {\bibfnamefont {J.}~\bibnamefont {Wambach}},\ }\bibfield  {title} {\bibinfo {title} {{The Phase Structure of the Polyakov--Quark-Meson Model}},\ }\href {https://doi.org/10.1103/PhysRevD.76.074023} {\bibfield  {journal} {\bibinfo  {journal} {Phys. Rev. D}\ }\textbf {\bibinfo {volume} {76}},\ \bibinfo {pages} {074023} (\bibinfo {year} {2007})},\ \Eprint {https://arxiv.org/abs/0704.3234} {arXiv:0704.3234 [hep-ph]} \BibitemShut {NoStop}%
\bibitem [{\citenamefont {Bhattacharyya}\ \emph {et~al.}(2013)\citenamefont {Bhattacharyya}, \citenamefont {Deb}, \citenamefont {Ghosh}, \citenamefont {Ray},\ and\ \citenamefont {Sur}}]{Bhattacharyya:2012rp}%
  \BibitemOpen
  \bibfield  {author} {\bibinfo {author} {\bibfnamefont {A.}~\bibnamefont {Bhattacharyya}}, \bibinfo {author} {\bibfnamefont {P.}~\bibnamefont {Deb}}, \bibinfo {author} {\bibfnamefont {S.~K.}\ \bibnamefont {Ghosh}}, \bibinfo {author} {\bibfnamefont {R.}~\bibnamefont {Ray}},\ and\ \bibinfo {author} {\bibfnamefont {S.}~\bibnamefont {Sur}},\ }\bibfield  {title} {\bibinfo {title} {{Thermodynamic Properties of Strongly Interacting Matter in Finite Volume using Polyakov-Nambu-Jona-Lasinio Model}},\ }\href {https://doi.org/10.1103/PhysRevD.87.054009} {\bibfield  {journal} {\bibinfo  {journal} {Phys. Rev. D}\ }\textbf {\bibinfo {volume} {87}},\ \bibinfo {pages} {054009} (\bibinfo {year} {2013})},\ \Eprint {https://arxiv.org/abs/1212.5893} {arXiv:1212.5893 [hep-ph]} \BibitemShut {NoStop}%
\bibitem [{\citenamefont {Wen}\ \emph {et~al.}(2019)\citenamefont {Wen}, \citenamefont {Huang},\ and\ \citenamefont {Fu}}]{Wen:2018nkn}%
  \BibitemOpen
  \bibfield  {author} {\bibinfo {author} {\bibfnamefont {R.}~\bibnamefont {Wen}}, \bibinfo {author} {\bibfnamefont {C.}~\bibnamefont {Huang}},\ and\ \bibinfo {author} {\bibfnamefont {W.-J.}\ \bibnamefont {Fu}},\ }\bibfield  {title} {\bibinfo {title} {{Baryon number fluctuations in the 2+1 flavor low energy effective model}},\ }\href {https://doi.org/10.1103/PhysRevD.99.094019} {\bibfield  {journal} {\bibinfo  {journal} {Phys. Rev. D}\ }\textbf {\bibinfo {volume} {99}},\ \bibinfo {pages} {094019} (\bibinfo {year} {2019})},\ \Eprint {https://arxiv.org/abs/1809.04233} {arXiv:1809.04233 [hep-ph]} \BibitemShut {NoStop}%
\bibitem [{\citenamefont {Wen}\ and\ \citenamefont {Fu}(2021)}]{Wen:2019ruz}%
  \BibitemOpen
  \bibfield  {author} {\bibinfo {author} {\bibfnamefont {R.}~\bibnamefont {Wen}}\ and\ \bibinfo {author} {\bibfnamefont {W.-j.}\ \bibnamefont {Fu}},\ }\bibfield  {title} {\bibinfo {title} {{Correlations of conserved charges and QCD phase structure}},\ }\href {https://doi.org/10.1088/1674-1137/abe199} {\bibfield  {journal} {\bibinfo  {journal} {Chin. Phys. C}\ }\textbf {\bibinfo {volume} {45}},\ \bibinfo {pages} {044112} (\bibinfo {year} {2021})},\ \Eprint {https://arxiv.org/abs/1909.12564} {arXiv:1909.12564 [hep-ph]} \BibitemShut {NoStop}%
\bibitem [{\citenamefont {Xu}\ and\ \citenamefont {Huang}(2013)}]{Xu:2011ud}%
  \BibitemOpen
  \bibfield  {author} {\bibinfo {author} {\bibfnamefont {F.}~\bibnamefont {Xu}}\ and\ \bibinfo {author} {\bibfnamefont {M.}~\bibnamefont {Huang}},\ }\bibfield  {title} {\bibinfo {title} {{Electric and magnetic screenings of gluons in a model with dimension-2 gluon condensate}},\ }\href {https://doi.org/10.1088/1674-1137/37/1/014103} {\bibfield  {journal} {\bibinfo  {journal} {Chin. Phys. C}\ }\textbf {\bibinfo {volume} {37}},\ \bibinfo {pages} {014103} (\bibinfo {year} {2013})},\ \Eprint {https://arxiv.org/abs/1111.5152} {arXiv:1111.5152 [hep-ph]} \BibitemShut {NoStop}%
\bibitem [{\citenamefont {Herbst}\ \emph {et~al.}(2014)\citenamefont {Herbst}, \citenamefont {Mitter}, \citenamefont {Pawlowski}, \citenamefont {Schaefer},\ and\ \citenamefont {Stiele}}]{Herbst:2013ufa}%
  \BibitemOpen
  \bibfield  {author} {\bibinfo {author} {\bibfnamefont {T.~K.}\ \bibnamefont {Herbst}}, \bibinfo {author} {\bibfnamefont {M.}~\bibnamefont {Mitter}}, \bibinfo {author} {\bibfnamefont {J.~M.}\ \bibnamefont {Pawlowski}}, \bibinfo {author} {\bibfnamefont {B.-J.}\ \bibnamefont {Schaefer}},\ and\ \bibinfo {author} {\bibfnamefont {R.}~\bibnamefont {Stiele}},\ }\bibfield  {title} {\bibinfo {title} {{Thermodynamics of QCD at vanishing density}},\ }\href {https://doi.org/10.1016/j.physletb.2014.02.045} {\bibfield  {journal} {\bibinfo  {journal} {Phys. Lett. B}\ }\textbf {\bibinfo {volume} {731}},\ \bibinfo {pages} {248} (\bibinfo {year} {2014})},\ \Eprint {https://arxiv.org/abs/1308.3621} {arXiv:1308.3621 [hep-ph]} \BibitemShut {NoStop}%
\bibitem [{\citenamefont {Gunkel}\ and\ \citenamefont {Fischer}(2021)}]{Gunkel:2021oya}%
  \BibitemOpen
  \bibfield  {author} {\bibinfo {author} {\bibfnamefont {P.~J.}\ \bibnamefont {Gunkel}}\ and\ \bibinfo {author} {\bibfnamefont {C.~S.}\ \bibnamefont {Fischer}},\ }\bibfield  {title} {\bibinfo {title} {{Locating the critical endpoint of QCD: Mesonic backcoupling effects}},\ }\href {https://doi.org/10.1103/PhysRevD.104.054022} {\bibfield  {journal} {\bibinfo  {journal} {Phys. Rev. D}\ }\textbf {\bibinfo {volume} {104}},\ \bibinfo {pages} {054022} (\bibinfo {year} {2021})},\ \Eprint {https://arxiv.org/abs/2106.08356} {arXiv:2106.08356 [hep-ph]} \BibitemShut {NoStop}%
\bibitem [{\citenamefont {Nakamura}\ \emph {et~al.}(2004)\citenamefont {Nakamura}, \citenamefont {Saito},\ and\ \citenamefont {Sakai}}]{Nakamura:2003pu}%
  \BibitemOpen
  \bibfield  {author} {\bibinfo {author} {\bibfnamefont {A.}~\bibnamefont {Nakamura}}, \bibinfo {author} {\bibfnamefont {T.}~\bibnamefont {Saito}},\ and\ \bibinfo {author} {\bibfnamefont {S.}~\bibnamefont {Sakai}},\ }\bibfield  {title} {\bibinfo {title} {{Lattice calculation of gluon screening masses}},\ }\href {https://doi.org/10.1103/PhysRevD.69.014506} {\bibfield  {journal} {\bibinfo  {journal} {Phys. Rev. D}\ }\textbf {\bibinfo {volume} {69}},\ \bibinfo {pages} {014506} (\bibinfo {year} {2004})},\ \Eprint {https://arxiv.org/abs/hep-lat/0311024} {arXiv:hep-lat/0311024} \BibitemShut {NoStop}%
\bibitem [{\citenamefont {Blaizot}\ \emph {et~al.}(2003)\citenamefont {Blaizot}, \citenamefont {Iancu},\ and\ \citenamefont {Rebhan}}]{Blaizot:2003tw}%
  \BibitemOpen
  \bibfield  {author} {\bibinfo {author} {\bibfnamefont {J.-P.}\ \bibnamefont {Blaizot}}, \bibinfo {author} {\bibfnamefont {E.}~\bibnamefont {Iancu}},\ and\ \bibinfo {author} {\bibfnamefont {A.}~\bibnamefont {Rebhan}},\ }\bibfield  {title} {\bibinfo {title} {{Thermodynamics of the high temperature quark gluon plasma}}\ }(\bibinfo {year} {2003})\ pp.\ \bibinfo {pages} {60--122},\ \Eprint {https://arxiv.org/abs/hep-ph/0303185} {arXiv:hep-ph/0303185} \BibitemShut {NoStop}%
\bibitem [{\citenamefont {Kraemmer}\ and\ \citenamefont {Rebhan}(2004)}]{Kraemmer:2003gd}%
  \BibitemOpen
  \bibfield  {author} {\bibinfo {author} {\bibfnamefont {U.}~\bibnamefont {Kraemmer}}\ and\ \bibinfo {author} {\bibfnamefont {A.}~\bibnamefont {Rebhan}},\ }\bibfield  {title} {\bibinfo {title} {{Advances in perturbative thermal field theory}},\ }\href {https://doi.org/10.1088/0034-4885/67/3/R05} {\bibfield  {journal} {\bibinfo  {journal} {Rept. Prog. Phys.}\ }\textbf {\bibinfo {volume} {67}},\ \bibinfo {pages} {351} (\bibinfo {year} {2004})},\ \Eprint {https://arxiv.org/abs/hep-ph/0310337} {arXiv:hep-ph/0310337} \BibitemShut {NoStop}%
\bibitem [{\citenamefont {Andersen}\ and\ \citenamefont {Strickland}(2005)}]{Andersen:2004fp}%
  \BibitemOpen
  \bibfield  {author} {\bibinfo {author} {\bibfnamefont {J.~O.}\ \bibnamefont {Andersen}}\ and\ \bibinfo {author} {\bibfnamefont {M.}~\bibnamefont {Strickland}},\ }\bibfield  {title} {\bibinfo {title} {{Resummation in hot field theories}},\ }\href {https://doi.org/10.1016/j.aop.2004.09.017} {\bibfield  {journal} {\bibinfo  {journal} {Annals Phys.}\ }\textbf {\bibinfo {volume} {317}},\ \bibinfo {pages} {281} (\bibinfo {year} {2005})},\ \Eprint {https://arxiv.org/abs/hep-ph/0404164} {arXiv:hep-ph/0404164} \BibitemShut {NoStop}%
\bibitem [{\citenamefont {Su}\ and\ \citenamefont {Tywoniuk}(2015)}]{Su:2014rma}%
  \BibitemOpen
  \bibfield  {author} {\bibinfo {author} {\bibfnamefont {N.}~\bibnamefont {Su}}\ and\ \bibinfo {author} {\bibfnamefont {K.}~\bibnamefont {Tywoniuk}},\ }\bibfield  {title} {\bibinfo {title} {{Massless Mode and Positivity Violation in Hot QCD}},\ }\href {https://doi.org/10.1103/PhysRevLett.114.161601} {\bibfield  {journal} {\bibinfo  {journal} {Phys. Rev. Lett.}\ }\textbf {\bibinfo {volume} {114}},\ \bibinfo {pages} {161601} (\bibinfo {year} {2015})},\ \Eprint {https://arxiv.org/abs/1409.3203} {arXiv:1409.3203 [hep-ph]} \BibitemShut {NoStop}%
\bibitem [{\citenamefont {Kapusta}\ and\ \citenamefont {Gale}(2011)}]{Kapusta:2006pm}%
  \BibitemOpen
  \bibfield  {author} {\bibinfo {author} {\bibfnamefont {J.~I.}\ \bibnamefont {Kapusta}}\ and\ \bibinfo {author} {\bibfnamefont {C.}~\bibnamefont {Gale}},\ }\href {https://doi.org/10.1017/CBO9780511535130} {\emph {\bibinfo {title} {{Finite-temperature field theory: Principles and applications}}}},\ Cambridge Monographs on Mathematical Physics\ (\bibinfo  {publisher} {Cambridge University Press},\ \bibinfo {year} {2011})\BibitemShut {NoStop}%
\bibitem [{\citenamefont {Schneider}\ and\ \citenamefont {Weise}(2001)}]{Schneider:2001nf}%
  \BibitemOpen
  \bibfield  {author} {\bibinfo {author} {\bibfnamefont {R.~A.}\ \bibnamefont {Schneider}}\ and\ \bibinfo {author} {\bibfnamefont {W.}~\bibnamefont {Weise}},\ }\bibfield  {title} {\bibinfo {title} {{On the quasiparticle description of lattice QCD thermodynamics}},\ }\href {https://doi.org/10.1103/PhysRevC.64.055201} {\bibfield  {journal} {\bibinfo  {journal} {Phys. Rev. C}\ }\textbf {\bibinfo {volume} {64}},\ \bibinfo {pages} {055201} (\bibinfo {year} {2001})},\ \Eprint {https://arxiv.org/abs/hep-ph/0105242} {arXiv:hep-ph/0105242} \BibitemShut {NoStop}%
\bibitem [{\citenamefont {Peshier}\ \emph {et~al.}(2002)\citenamefont {Peshier}, \citenamefont {Kampfer},\ and\ \citenamefont {Soff}}]{Peshier:2002ww}%
  \BibitemOpen
  \bibfield  {author} {\bibinfo {author} {\bibfnamefont {A.}~\bibnamefont {Peshier}}, \bibinfo {author} {\bibfnamefont {B.}~\bibnamefont {Kampfer}},\ and\ \bibinfo {author} {\bibfnamefont {G.}~\bibnamefont {Soff}},\ }\bibfield  {title} {\bibinfo {title} {{From QCD lattice calculations to the equation of state of quark matter}},\ }\href {https://doi.org/10.1103/PhysRevD.66.094003} {\bibfield  {journal} {\bibinfo  {journal} {Phys. Rev. D}\ }\textbf {\bibinfo {volume} {66}},\ \bibinfo {pages} {094003} (\bibinfo {year} {2002})},\ \Eprint {https://arxiv.org/abs/hep-ph/0206229} {arXiv:hep-ph/0206229} \BibitemShut {NoStop}%
\bibitem [{\citenamefont {Bluhm}\ \emph {et~al.}(2009)\citenamefont {Bluhm}, \citenamefont {Kampfer}, \citenamefont {Schulze},\ and\ \citenamefont {Seipt}}]{Bluhm:2009wd}%
  \BibitemOpen
  \bibfield  {author} {\bibinfo {author} {\bibfnamefont {M.}~\bibnamefont {Bluhm}}, \bibinfo {author} {\bibfnamefont {B.}~\bibnamefont {Kampfer}}, \bibinfo {author} {\bibfnamefont {R.}~\bibnamefont {Schulze}},\ and\ \bibinfo {author} {\bibfnamefont {D.}~\bibnamefont {Seipt}},\ }\bibfield  {title} {\bibinfo {title} {{QCD equation of state: Physical quark masses and asymptotic temperatures}},\ }\href {https://doi.org/10.1016/j.ppnp.2008.12.041} {\bibfield  {journal} {\bibinfo  {journal} {Prog. Part. Nucl. Phys.}\ }\textbf {\bibinfo {volume} {62}},\ \bibinfo {pages} {512} (\bibinfo {year} {2009})},\ \Eprint {https://arxiv.org/abs/0901.0472} {arXiv:0901.0472 [hep-ph]} \BibitemShut {NoStop}%
\bibitem [{\citenamefont {Alba}\ \emph {et~al.}(2014)\citenamefont {Alba}, \citenamefont {Alberico}, \citenamefont {Bluhm}, \citenamefont {Greco}, \citenamefont {Ratti},\ and\ \citenamefont {Ruggieri}}]{Alba:2014lda}%
  \BibitemOpen
  \bibfield  {author} {\bibinfo {author} {\bibfnamefont {P.}~\bibnamefont {Alba}}, \bibinfo {author} {\bibfnamefont {W.}~\bibnamefont {Alberico}}, \bibinfo {author} {\bibfnamefont {M.}~\bibnamefont {Bluhm}}, \bibinfo {author} {\bibfnamefont {V.}~\bibnamefont {Greco}}, \bibinfo {author} {\bibfnamefont {C.}~\bibnamefont {Ratti}},\ and\ \bibinfo {author} {\bibfnamefont {M.}~\bibnamefont {Ruggieri}},\ }\bibfield  {title} {\bibinfo {title} {{Polyakov loop and gluon quasiparticles: A self-consistent approach to Yang{\textendash}Mills thermodynamics}},\ }\href {https://doi.org/10.1016/j.nuclphysa.2014.11.011} {\bibfield  {journal} {\bibinfo  {journal} {Nucl. Phys. A}\ }\textbf {\bibinfo {volume} {934}},\ \bibinfo {pages} {41} (\bibinfo {year} {2014})},\ \Eprint {https://arxiv.org/abs/1402.6213} {arXiv:1402.6213 [hep-ph]} \BibitemShut {NoStop}%
\bibitem [{\citenamefont {Koothottil}\ and\ \citenamefont {Bannur}(2019)}]{Koothottil:2018akg}%
  \BibitemOpen
  \bibfield  {author} {\bibinfo {author} {\bibfnamefont {S.}~\bibnamefont {Koothottil}}\ and\ \bibinfo {author} {\bibfnamefont {V.~M.}\ \bibnamefont {Bannur}},\ }\bibfield  {title} {\bibinfo {title} {{Thermodynamic behavior of magnetized quark-gluon plasma within the self-consistent quasiparticle model}},\ }\href {https://doi.org/10.1103/PhysRevC.99.035210} {\bibfield  {journal} {\bibinfo  {journal} {Phys. Rev. C}\ }\textbf {\bibinfo {volume} {99}},\ \bibinfo {pages} {035210} (\bibinfo {year} {2019})},\ \Eprint {https://arxiv.org/abs/1811.05377} {arXiv:1811.05377 [nucl-th]} \BibitemShut {NoStop}%
\bibitem [{\citenamefont {Cybenko}(1989)}]{cybenko1989approximation}%
  \BibitemOpen
  \bibfield  {author} {\bibinfo {author} {\bibfnamefont {G.}~\bibnamefont {Cybenko}},\ }\bibfield  {title} {\bibinfo {title} {Approximation by superpositions of a sigmoidal function},\ }\href@noop {} {\bibfield  {journal} {\bibinfo  {journal} {Mathematics of control, signals and systems}\ }\textbf {\bibinfo {volume} {2}},\ \bibinfo {pages} {303} (\bibinfo {year} {1989})}\BibitemShut {NoStop}%
\bibitem [{\citenamefont {Hornik}\ \emph {et~al.}(1989)\citenamefont {Hornik}, \citenamefont {Stinchcombe},\ and\ \citenamefont {White}}]{hornik1989multilayer}%
  \BibitemOpen
  \bibfield  {author} {\bibinfo {author} {\bibfnamefont {K.}~\bibnamefont {Hornik}}, \bibinfo {author} {\bibfnamefont {M.}~\bibnamefont {Stinchcombe}},\ and\ \bibinfo {author} {\bibfnamefont {H.}~\bibnamefont {White}},\ }\bibfield  {title} {\bibinfo {title} {Multilayer feedforward networks are universal approximators},\ }\href@noop {} {\bibfield  {journal} {\bibinfo  {journal} {Neural networks}\ }\textbf {\bibinfo {volume} {2}},\ \bibinfo {pages} {359} (\bibinfo {year} {1989})}\BibitemShut {NoStop}%
\bibitem [{\citenamefont {Raissi}\ \emph {et~al.}(2019)\citenamefont {Raissi}, \citenamefont {Perdikaris},\ and\ \citenamefont {Karniadakis}}]{raissi2019physics}%
  \BibitemOpen
  \bibfield  {author} {\bibinfo {author} {\bibfnamefont {M.}~\bibnamefont {Raissi}}, \bibinfo {author} {\bibfnamefont {P.}~\bibnamefont {Perdikaris}},\ and\ \bibinfo {author} {\bibfnamefont {G.~E.}\ \bibnamefont {Karniadakis}},\ }\bibfield  {title} {\bibinfo {title} {Physics-informed neural networks: A deep learning framework for solving forward and inverse problems involving nonlinear partial differential equations},\ }\href@noop {} {\bibfield  {journal} {\bibinfo  {journal} {Journal of Computational physics}\ }\textbf {\bibinfo {volume} {378}},\ \bibinfo {pages} {686} (\bibinfo {year} {2019})}\BibitemShut {NoStop}%
\bibitem [{\citenamefont {Aarts}\ \emph {et~al.}(2025)\citenamefont {Aarts}, \citenamefont {Fukushima}, \citenamefont {Hatsuda}, \citenamefont {Ipp}, \citenamefont {Shi}, \citenamefont {Wang},\ and\ \citenamefont {Zhou}}]{DL_nuclear}%
  \BibitemOpen
  \bibfield  {author} {\bibinfo {author} {\bibfnamefont {G.}~\bibnamefont {Aarts}}, \bibinfo {author} {\bibfnamefont {K.}~\bibnamefont {Fukushima}}, \bibinfo {author} {\bibfnamefont {T.}~\bibnamefont {Hatsuda}}, \bibinfo {author} {\bibfnamefont {A.}~\bibnamefont {Ipp}}, \bibinfo {author} {\bibfnamefont {S.}~\bibnamefont {Shi}}, \bibinfo {author} {\bibfnamefont {L.}~\bibnamefont {Wang}},\ and\ \bibinfo {author} {\bibfnamefont {K.}~\bibnamefont {Zhou}},\ }\bibfield  {title} {\bibinfo {title} {{Physics-driven learning for inverse problems in quantum chromodynamics}},\ }\href {https://doi.org/10.1038/s42254-024-00798-x} {\bibfield  {journal} {\bibinfo  {journal} {Nature Rev. Phys.}\ }\textbf {\bibinfo {volume} {7}},\ \bibinfo {pages} {154} (\bibinfo {year} {2025})},\ \Eprint {https://arxiv.org/abs/2501.05580} {arXiv:2501.05580 [hep-lat]} \BibitemShut {NoStop}%
\bibitem [{\citenamefont {Pang}\ \emph {et~al.}(2018)\citenamefont {Pang}, \citenamefont {Zhou}, \citenamefont {Su}, \citenamefont {Petersen}, \citenamefont {St{\"o}cker},\ and\ \citenamefont {Wang}}]{DL_nuclear1}%
  \BibitemOpen
  \bibfield  {author} {\bibinfo {author} {\bibfnamefont {L.-G.}\ \bibnamefont {Pang}}, \bibinfo {author} {\bibfnamefont {K.}~\bibnamefont {Zhou}}, \bibinfo {author} {\bibfnamefont {N.}~\bibnamefont {Su}}, \bibinfo {author} {\bibfnamefont {H.}~\bibnamefont {Petersen}}, \bibinfo {author} {\bibfnamefont {H.}~\bibnamefont {St{\"o}cker}},\ and\ \bibinfo {author} {\bibfnamefont {X.-N.}\ \bibnamefont {Wang}},\ }\bibfield  {title} {\bibinfo {title} {An equation-of-state-meter of quantum chromodynamics transition from deep learning},\ }\href@noop {} {\bibfield  {journal} {\bibinfo  {journal} {Nature communications}\ }\textbf {\bibinfo {volume} {9}},\ \bibinfo {pages} {210} (\bibinfo {year} {2018})}\BibitemShut {NoStop}%
\bibitem [{\citenamefont {Liu}\ \emph {et~al.}(2019)\citenamefont {Liu}, \citenamefont {Zhao},\ and\ \citenamefont {Song}}]{DL_nuclear2}%
  \BibitemOpen
  \bibfield  {author} {\bibinfo {author} {\bibfnamefont {Z.}~\bibnamefont {Liu}}, \bibinfo {author} {\bibfnamefont {W.}~\bibnamefont {Zhao}},\ and\ \bibinfo {author} {\bibfnamefont {H.}~\bibnamefont {Song}},\ }\bibfield  {title} {\bibinfo {title} {Principal component analysis of collective flow in relativistic heavy-ion collisions},\ }\href@noop {} {\bibfield  {journal} {\bibinfo  {journal} {The European Physical Journal C}\ }\textbf {\bibinfo {volume} {79}},\ \bibinfo {pages} {870} (\bibinfo {year} {2019})}\BibitemShut {NoStop}%
\bibitem [{\citenamefont {Huang}\ \emph {et~al.}(2021)\citenamefont {Huang}, \citenamefont {Xiao}, \citenamefont {Liu}, \citenamefont {Wu}, \citenamefont {Mu},\ and\ \citenamefont {Song}}]{DL_nuclear3}%
  \BibitemOpen
  \bibfield  {author} {\bibinfo {author} {\bibfnamefont {H.}~\bibnamefont {Huang}}, \bibinfo {author} {\bibfnamefont {B.}~\bibnamefont {Xiao}}, \bibinfo {author} {\bibfnamefont {Z.}~\bibnamefont {Liu}}, \bibinfo {author} {\bibfnamefont {Z.}~\bibnamefont {Wu}}, \bibinfo {author} {\bibfnamefont {Y.}~\bibnamefont {Mu}},\ and\ \bibinfo {author} {\bibfnamefont {H.}~\bibnamefont {Song}},\ }\bibfield  {title} {\bibinfo {title} {Applications of deep learning to relativistic hydrodynamics},\ }\href@noop {} {\bibfield  {journal} {\bibinfo  {journal} {Physical Review Research}\ }\textbf {\bibinfo {volume} {3}},\ \bibinfo {pages} {023256} (\bibinfo {year} {2021})}\BibitemShut {NoStop}%
\bibitem [{\citenamefont {Shi}\ \emph {et~al.}(2022)\citenamefont {Shi}, \citenamefont {Zhou}, \citenamefont {Zhao}, \citenamefont {Mukherjee},\ and\ \citenamefont {Zhuang}}]{DL_nuclear4}%
  \BibitemOpen
  \bibfield  {author} {\bibinfo {author} {\bibfnamefont {S.}~\bibnamefont {Shi}}, \bibinfo {author} {\bibfnamefont {K.}~\bibnamefont {Zhou}}, \bibinfo {author} {\bibfnamefont {J.}~\bibnamefont {Zhao}}, \bibinfo {author} {\bibfnamefont {S.}~\bibnamefont {Mukherjee}},\ and\ \bibinfo {author} {\bibfnamefont {P.}~\bibnamefont {Zhuang}},\ }\bibfield  {title} {\bibinfo {title} {Heavy quark potential in the quark-gluon plasma: Deep neural network meets lattice quantum chromodynamics},\ }\href@noop {} {\bibfield  {journal} {\bibinfo  {journal} {Physical Review D}\ }\textbf {\bibinfo {volume} {105}},\ \bibinfo {pages} {014017} (\bibinfo {year} {2022})}\BibitemShut {NoStop}%
\bibitem [{\citenamefont {Wang}\ \emph {et~al.}(2022)\citenamefont {Wang}, \citenamefont {Shi},\ and\ \citenamefont {Zhou}}]{DL_nuclear5}%
  \BibitemOpen
  \bibfield  {author} {\bibinfo {author} {\bibfnamefont {L.}~\bibnamefont {Wang}}, \bibinfo {author} {\bibfnamefont {S.}~\bibnamefont {Shi}},\ and\ \bibinfo {author} {\bibfnamefont {K.}~\bibnamefont {Zhou}},\ }\bibfield  {title} {\bibinfo {title} {Reconstructing spectral functions via automatic differentiation},\ }\href@noop {} {\bibfield  {journal} {\bibinfo  {journal} {Physical Review D}\ }\textbf {\bibinfo {volume} {106}},\ \bibinfo {pages} {L051502} (\bibinfo {year} {2022})}\BibitemShut {NoStop}%
\bibitem [{\citenamefont {Wang}(2025)}]{DL_nuclear6}%
  \BibitemOpen
  \bibfield  {author} {\bibinfo {author} {\bibfnamefont {L.}~\bibnamefont {Wang}},\ }\bibfield  {title} {\bibinfo {title} {{Deep learning for exploring hadron{\textendash}hadron interactions}},\ }\href {https://doi.org/10.1016/j.jspc.2025.100024} {\bibfield  {journal} {\bibinfo  {journal} {J. Subatomic Part. Cosmol.}\ }\textbf {\bibinfo {volume} {3}},\ \bibinfo {pages} {100024} (\bibinfo {year} {2025})},\ \Eprint {https://arxiv.org/abs/2501.00374} {arXiv:2501.00374 [nucl-th]} \BibitemShut {NoStop}%
\bibitem [{\citenamefont {Wang}\ and\ \citenamefont {Zhao}(2024)}]{DL_nuclear7}%
  \BibitemOpen
  \bibfield  {author} {\bibinfo {author} {\bibfnamefont {L.}~\bibnamefont {Wang}}\ and\ \bibinfo {author} {\bibfnamefont {J.}~\bibnamefont {Zhao}},\ }\href@noop {} {\bibinfo {title} {{Learning Hadron Emitting Sources with Deep Neural Networks}}} (\bibinfo {year} {2024}),\ \Eprint {https://arxiv.org/abs/2411.16343} {arXiv:2411.16343 [nucl-th]} \BibitemShut {NoStop}%
\bibitem [{\citenamefont {Wang}\ \emph {et~al.}(2025)\citenamefont {Wang}, \citenamefont {Doi}, \citenamefont {Hatsuda},\ and\ \citenamefont {Lyu}}]{DL_nuclear8}%
  \BibitemOpen
  \bibfield  {author} {\bibinfo {author} {\bibfnamefont {L.}~\bibnamefont {Wang}}, \bibinfo {author} {\bibfnamefont {T.}~\bibnamefont {Doi}}, \bibinfo {author} {\bibfnamefont {T.}~\bibnamefont {Hatsuda}},\ and\ \bibinfo {author} {\bibfnamefont {Y.}~\bibnamefont {Lyu}},\ }\bibfield  {title} {\bibinfo {title} {{Building Hadron Potentials from Lattice QCD with Deep Neural Networks}},\ }\href {https://doi.org/10.22323/1.466.0076} {\bibfield  {journal} {\bibinfo  {journal} {PoS}\ }\textbf {\bibinfo {volume} {LATTICE2024}},\ \bibinfo {pages} {076} (\bibinfo {year} {2025})},\ \Eprint {https://arxiv.org/abs/2410.03082} {arXiv:2410.03082 [hep-lat]} \BibitemShut {NoStop}%
\bibitem [{\citenamefont {Zhou}\ \emph {et~al.}(2024)\citenamefont {Zhou}, \citenamefont {Wang}, \citenamefont {Pang},\ and\ \citenamefont {Shi}}]{Zhou:2023pti}%
  \BibitemOpen
  \bibfield  {author} {\bibinfo {author} {\bibfnamefont {K.}~\bibnamefont {Zhou}}, \bibinfo {author} {\bibfnamefont {L.}~\bibnamefont {Wang}}, \bibinfo {author} {\bibfnamefont {L.-G.}\ \bibnamefont {Pang}},\ and\ \bibinfo {author} {\bibfnamefont {S.}~\bibnamefont {Shi}},\ }\bibfield  {title} {\bibinfo {title} {{Exploring QCD matter in extreme conditions with Machine Learning}},\ }\href {https://doi.org/10.1016/j.ppnp.2023.104084} {\bibfield  {journal} {\bibinfo  {journal} {Prog. Part. Nucl. Phys.}\ }\textbf {\bibinfo {volume} {135}},\ \bibinfo {pages} {104084} (\bibinfo {year} {2024})},\ \Eprint {https://arxiv.org/abs/2303.15136} {arXiv:2303.15136 [hep-ph]} \BibitemShut {NoStop}%
\bibitem [{\citenamefont {Chen}\ \emph {et~al.}(2025)\citenamefont {Chen}, \citenamefont {Chen},\ and\ \citenamefont {Zhou}}]{Chen:2025kqb}%
  \BibitemOpen
  \bibfield  {author} {\bibinfo {author} {\bibfnamefont {X.}~\bibnamefont {Chen}}, \bibinfo {author} {\bibfnamefont {Y.}~\bibnamefont {Chen}},\ and\ \bibinfo {author} {\bibfnamefont {K.}~\bibnamefont {Zhou}},\ }\bibfield  {title} {\bibinfo {title} {{Data-Driven Einstein-Dilaton Model for Pure Yang-Mills Thermodynamics and Glueball Spectrum}},\ }\href@noop {} {\  (\bibinfo {year} {2025})},\ \Eprint {https://arxiv.org/abs/2507.06729} {arXiv:2507.06729 [hep-ph]} \BibitemShut {NoStop}%
\bibitem [{\citenamefont {Dai}\ \emph {et~al.}(2025)\citenamefont {Dai}, \citenamefont {Luo}, \citenamefont {Chen}, \citenamefont {Chen}, \citenamefont {Zhu},\ and\ \citenamefont {Li}}]{Dai:2025dir}%
  \BibitemOpen
  \bibfield  {author} {\bibinfo {author} {\bibfnamefont {W.-C.}\ \bibnamefont {Dai}}, \bibinfo {author} {\bibfnamefont {O.-Y.}\ \bibnamefont {Luo}}, \bibinfo {author} {\bibfnamefont {B.}~\bibnamefont {Chen}}, \bibinfo {author} {\bibfnamefont {X.}~\bibnamefont {Chen}}, \bibinfo {author} {\bibfnamefont {X.-Y.}\ \bibnamefont {Zhu}},\ and\ \bibinfo {author} {\bibfnamefont {X.-H.}\ \bibnamefont {Li}},\ }\bibfield  {title} {\bibinfo {title} {{Extracting Transport Properties of Quark-Gluon Plasma from the Heavy-Quark Potential With Neural Networks in a Holographic Model}},\ }\href@noop {} {\  (\bibinfo {year} {2025})},\ \Eprint {https://arxiv.org/abs/2503.10213} {arXiv:2503.10213 [hep-ph]} \BibitemShut {NoStop}%
\bibitem [{\citenamefont {Mansouri}\ \emph {et~al.}(2024)\citenamefont {Mansouri}, \citenamefont {Bitaghsir~Fadafan},\ and\ \citenamefont {Chen}}]{Mansouri:2024uwc}%
  \BibitemOpen
  \bibfield  {author} {\bibinfo {author} {\bibfnamefont {M.}~\bibnamefont {Mansouri}}, \bibinfo {author} {\bibfnamefont {K.}~\bibnamefont {Bitaghsir~Fadafan}},\ and\ \bibinfo {author} {\bibfnamefont {X.}~\bibnamefont {Chen}},\ }\bibfield  {title} {\bibinfo {title} {{Holographic complex potential of a quarkonium from deep learning}},\ }\href@noop {} {\  (\bibinfo {year} {2024})},\ \Eprint {https://arxiv.org/abs/2406.06285} {arXiv:2406.06285 [hep-ph]} \BibitemShut {NoStop}%
\bibitem [{\citenamefont {Baldi}\ \emph {et~al.}(2014)\citenamefont {Baldi}, \citenamefont {Sadowski},\ and\ \citenamefont {Whiteson}}]{DL_particle1}%
  \BibitemOpen
  \bibfield  {author} {\bibinfo {author} {\bibfnamefont {P.}~\bibnamefont {Baldi}}, \bibinfo {author} {\bibfnamefont {P.}~\bibnamefont {Sadowski}},\ and\ \bibinfo {author} {\bibfnamefont {D.}~\bibnamefont {Whiteson}},\ }\bibfield  {title} {\bibinfo {title} {Searching for exotic particles in high-energy physics with deep learning},\ }\href@noop {} {\bibfield  {journal} {\bibinfo  {journal} {Nature communications}\ }\textbf {\bibinfo {volume} {5}},\ \bibinfo {pages} {4308} (\bibinfo {year} {2014})}\BibitemShut {NoStop}%
\bibitem [{\citenamefont {Baldi}\ \emph {et~al.}(2015)\citenamefont {Baldi}, \citenamefont {Sadowski},\ and\ \citenamefont {Whiteson}}]{DL_particle2}%
  \BibitemOpen
  \bibfield  {author} {\bibinfo {author} {\bibfnamefont {P.}~\bibnamefont {Baldi}}, \bibinfo {author} {\bibfnamefont {P.}~\bibnamefont {Sadowski}},\ and\ \bibinfo {author} {\bibfnamefont {D.}~\bibnamefont {Whiteson}},\ }\bibfield  {title} {\bibinfo {title} {Enhanced higgs boson to $\tau$+ $\tau$-search with deep learning},\ }\href@noop {} {\bibfield  {journal} {\bibinfo  {journal} {Physical review letters}\ }\textbf {\bibinfo {volume} {114}},\ \bibinfo {pages} {111801} (\bibinfo {year} {2015})}\BibitemShut {NoStop}%
\bibitem [{\citenamefont {Barnard}\ \emph {et~al.}(2017)\citenamefont {Barnard}, \citenamefont {Dawe}, \citenamefont {Dolan},\ and\ \citenamefont {Rajcic}}]{DL_particle3}%
  \BibitemOpen
  \bibfield  {author} {\bibinfo {author} {\bibfnamefont {J.}~\bibnamefont {Barnard}}, \bibinfo {author} {\bibfnamefont {E.~N.}\ \bibnamefont {Dawe}}, \bibinfo {author} {\bibfnamefont {M.~J.}\ \bibnamefont {Dolan}},\ and\ \bibinfo {author} {\bibfnamefont {N.}~\bibnamefont {Rajcic}},\ }\bibfield  {title} {\bibinfo {title} {Parton shower uncertainties in jet substructure analyses with deep neural networks},\ }\href@noop {} {\bibfield  {journal} {\bibinfo  {journal} {Physical Review D}\ }\textbf {\bibinfo {volume} {95}},\ \bibinfo {pages} {014018} (\bibinfo {year} {2017})}\BibitemShut {NoStop}%
\bibitem [{\citenamefont {Broecker}\ \emph {et~al.}(2017)\citenamefont {Broecker}, \citenamefont {Carrasquilla}, \citenamefont {Melko},\ and\ \citenamefont {Trebst}}]{DL_particle4}%
  \BibitemOpen
  \bibfield  {author} {\bibinfo {author} {\bibfnamefont {P.}~\bibnamefont {Broecker}}, \bibinfo {author} {\bibfnamefont {J.}~\bibnamefont {Carrasquilla}}, \bibinfo {author} {\bibfnamefont {R.~G.}\ \bibnamefont {Melko}},\ and\ \bibinfo {author} {\bibfnamefont {S.}~\bibnamefont {Trebst}},\ }\bibfield  {title} {\bibinfo {title} {Machine learning quantum phases of matter beyond the fermion sign problem},\ }\href@noop {} {\bibfield  {journal} {\bibinfo  {journal} {Scientific reports}\ }\textbf {\bibinfo {volume} {7}},\ \bibinfo {pages} {8823} (\bibinfo {year} {2017})}\BibitemShut {NoStop}%
\bibitem [{\citenamefont {Radovic}\ \emph {et~al.}(2018)\citenamefont {Radovic}, \citenamefont {Williams}, \citenamefont {Rousseau}, \citenamefont {Kagan}, \citenamefont {Bonacorsi}, \citenamefont {Himmel}, \citenamefont {Aurisano}, \citenamefont {Terao},\ and\ \citenamefont {Wongjirad}}]{DL_particle5}%
  \BibitemOpen
  \bibfield  {author} {\bibinfo {author} {\bibfnamefont {A.}~\bibnamefont {Radovic}}, \bibinfo {author} {\bibfnamefont {M.}~\bibnamefont {Williams}}, \bibinfo {author} {\bibfnamefont {D.}~\bibnamefont {Rousseau}}, \bibinfo {author} {\bibfnamefont {M.}~\bibnamefont {Kagan}}, \bibinfo {author} {\bibfnamefont {D.}~\bibnamefont {Bonacorsi}}, \bibinfo {author} {\bibfnamefont {A.}~\bibnamefont {Himmel}}, \bibinfo {author} {\bibfnamefont {A.}~\bibnamefont {Aurisano}}, \bibinfo {author} {\bibfnamefont {K.}~\bibnamefont {Terao}},\ and\ \bibinfo {author} {\bibfnamefont {T.}~\bibnamefont {Wongjirad}},\ }\bibfield  {title} {\bibinfo {title} {Machine learning at the energy and intensity frontiers of particle physics},\ }\href@noop {} {\bibfield  {journal} {\bibinfo  {journal} {Nature}\ }\textbf {\bibinfo {volume} {560}},\ \bibinfo {pages} {41} (\bibinfo {year} {2018})}\BibitemShut {NoStop}%
\bibitem [{\citenamefont {Wang}\ \emph {et~al.}(2024)\citenamefont {Wang}, \citenamefont {Aarts},\ and\ \citenamefont {Zhou}}]{Wang:2023exq}%
  \BibitemOpen
  \bibfield  {author} {\bibinfo {author} {\bibfnamefont {L.}~\bibnamefont {Wang}}, \bibinfo {author} {\bibfnamefont {G.}~\bibnamefont {Aarts}},\ and\ \bibinfo {author} {\bibfnamefont {K.}~\bibnamefont {Zhou}},\ }\bibfield  {title} {\bibinfo {title} {{Diffusion models as stochastic quantization in lattice field theory}},\ }\href {https://doi.org/10.1007/JHEP05(2024)060} {\bibfield  {journal} {\bibinfo  {journal} {JHEP}\ }\textbf {\bibinfo {volume} {05}},\ \bibinfo {pages} {060}},\ \Eprint {https://arxiv.org/abs/2309.17082} {arXiv:2309.17082 [hep-lat]} \BibitemShut {NoStop}%
\bibitem [{\citenamefont {Carrasquilla}\ and\ \citenamefont {Melko}(2017)}]{DL_condensed1}%
  \BibitemOpen
  \bibfield  {author} {\bibinfo {author} {\bibfnamefont {J.}~\bibnamefont {Carrasquilla}}\ and\ \bibinfo {author} {\bibfnamefont {R.~G.}\ \bibnamefont {Melko}},\ }\bibfield  {title} {\bibinfo {title} {Machine learning phases of matter},\ }\href@noop {} {\bibfield  {journal} {\bibinfo  {journal} {Nature Physics}\ }\textbf {\bibinfo {volume} {13}},\ \bibinfo {pages} {431} (\bibinfo {year} {2017})}\BibitemShut {NoStop}%
\bibitem [{\citenamefont {Carleo}\ \emph {et~al.}(2019)\citenamefont {Carleo}, \citenamefont {Cirac}, \citenamefont {Cranmer}, \citenamefont {Daudet}, \citenamefont {Schuld}, \citenamefont {Tishby}, \citenamefont {Vogt-Maranto},\ and\ \citenamefont {Zdeborov{\'a}}}]{DL_condensed2}%
  \BibitemOpen
  \bibfield  {author} {\bibinfo {author} {\bibfnamefont {G.}~\bibnamefont {Carleo}}, \bibinfo {author} {\bibfnamefont {I.}~\bibnamefont {Cirac}}, \bibinfo {author} {\bibfnamefont {K.}~\bibnamefont {Cranmer}}, \bibinfo {author} {\bibfnamefont {L.}~\bibnamefont {Daudet}}, \bibinfo {author} {\bibfnamefont {M.}~\bibnamefont {Schuld}}, \bibinfo {author} {\bibfnamefont {N.}~\bibnamefont {Tishby}}, \bibinfo {author} {\bibfnamefont {L.}~\bibnamefont {Vogt-Maranto}},\ and\ \bibinfo {author} {\bibfnamefont {L.}~\bibnamefont {Zdeborov{\'a}}},\ }\bibfield  {title} {\bibinfo {title} {Machine learning and the physical sciences},\ }\href@noop {} {\bibfield  {journal} {\bibinfo  {journal} {Reviews of Modern Physics}\ }\textbf {\bibinfo {volume} {91}},\ \bibinfo {pages} {045002} (\bibinfo {year} {2019})}\BibitemShut {NoStop}%
\bibitem [{\citenamefont {Li}\ \emph {et~al.}(2023)\citenamefont {Li}, \citenamefont {L{\"u}}, \citenamefont {Pang},\ and\ \citenamefont {Qin}}]{Li:2022ozl}%
  \BibitemOpen
  \bibfield  {author} {\bibinfo {author} {\bibfnamefont {F.-P.}\ \bibnamefont {Li}}, \bibinfo {author} {\bibfnamefont {H.-L.}\ \bibnamefont {L{\"u}}}, \bibinfo {author} {\bibfnamefont {L.-G.}\ \bibnamefont {Pang}},\ and\ \bibinfo {author} {\bibfnamefont {G.-Y.}\ \bibnamefont {Qin}},\ }\bibfield  {title} {\bibinfo {title} {{Deep-learning quasi-particle masses from QCD equation of state}},\ }\href {https://doi.org/10.1016/j.physletb.2023.138088} {\bibfield  {journal} {\bibinfo  {journal} {Phys. Lett. B}\ }\textbf {\bibinfo {volume} {844}},\ \bibinfo {pages} {138088} (\bibinfo {year} {2023})},\ \Eprint {https://arxiv.org/abs/2211.07994} {arXiv:2211.07994 [hep-ph]} \BibitemShut {NoStop}%
\bibitem [{\citenamefont {Li}\ \emph {et~al.}(2025)\citenamefont {Li}, \citenamefont {Pang},\ and\ \citenamefont {Qin}}]{Li:2025csc}%
  \BibitemOpen
  \bibfield  {author} {\bibinfo {author} {\bibfnamefont {F.-P.}\ \bibnamefont {Li}}, \bibinfo {author} {\bibfnamefont {L.-G.}\ \bibnamefont {Pang}},\ and\ \bibinfo {author} {\bibfnamefont {G.-Y.}\ \bibnamefont {Qin}},\ }\bibfield  {title} {\bibinfo {title} {{QCD equation of state at finite {\ensuremath{\mu}}B using deep learning assisted quasi-parton model}},\ }\href {https://doi.org/10.1016/j.physletb.2025.139692} {\bibfield  {journal} {\bibinfo  {journal} {Phys. Lett. B}\ }\textbf {\bibinfo {volume} {868}},\ \bibinfo {pages} {139692} (\bibinfo {year} {2025})},\ \Eprint {https://arxiv.org/abs/2501.10012} {arXiv:2501.10012 [nucl-th]} \BibitemShut {NoStop}%
\bibitem [{\citenamefont {He}\ \emph {et~al.}(2016)\citenamefont {He}, \citenamefont {Zhang}, \citenamefont {Ren},\ and\ \citenamefont {Sun}}]{he2016deep}%
  \BibitemOpen
  \bibfield  {author} {\bibinfo {author} {\bibfnamefont {K.}~\bibnamefont {He}}, \bibinfo {author} {\bibfnamefont {X.}~\bibnamefont {Zhang}}, \bibinfo {author} {\bibfnamefont {S.}~\bibnamefont {Ren}},\ and\ \bibinfo {author} {\bibfnamefont {J.}~\bibnamefont {Sun}},\ }\bibfield  {title} {\bibinfo {title} {Deep residual learning for image recognition},\ }in\ \href {https://doi.org/10.1109/CVPR.2016.90} {\emph {\bibinfo {booktitle} {2016 IEEE Conference on CVPR}}}\ (\bibinfo {year} {2016})\ pp.\ \bibinfo {pages} {770--778}\BibitemShut {NoStop}%
\bibitem [{\citenamefont {Yu}\ \emph {et~al.}(2018)\citenamefont {Yu}, \citenamefont {Yu},\ and\ \citenamefont {Ramalingam}}]{he2016identity}%
  \BibitemOpen
  \bibfield  {author} {\bibinfo {author} {\bibfnamefont {X.}~\bibnamefont {Yu}}, \bibinfo {author} {\bibfnamefont {Z.}~\bibnamefont {Yu}},\ and\ \bibinfo {author} {\bibfnamefont {S.}~\bibnamefont {Ramalingam}},\ }\bibfield  {title} {\bibinfo {title} {Learning strict identity mappings in deep residual networks},\ }in\ \href {https://doi.org/10.1109/CVPR.2018.00466} {\emph {\bibinfo {booktitle} {2018 IEEE Conference on CVPR}}}\ (\bibinfo {year} {2018})\ pp.\ \bibinfo {pages} {4432--4440}\BibitemShut {NoStop}%
\bibitem [{\citenamefont {Borsanyi}\ \emph {et~al.}(2012)\citenamefont {Borsanyi}, \citenamefont {Endrodi}, \citenamefont {Fodor}, \citenamefont {Katz},\ and\ \citenamefont {Szabo}}]{Borsanyi:2012ve}%
  \BibitemOpen
  \bibfield  {author} {\bibinfo {author} {\bibfnamefont {S.}~\bibnamefont {Borsanyi}}, \bibinfo {author} {\bibfnamefont {G.}~\bibnamefont {Endrodi}}, \bibinfo {author} {\bibfnamefont {Z.}~\bibnamefont {Fodor}}, \bibinfo {author} {\bibfnamefont {S.~D.}\ \bibnamefont {Katz}},\ and\ \bibinfo {author} {\bibfnamefont {K.~K.}\ \bibnamefont {Szabo}},\ }\bibfield  {title} {\bibinfo {title} {{Precision SU(3) lattice thermodynamics for a large temperature range}},\ }\href {https://doi.org/10.1007/JHEP07(2012)056} {\bibfield  {journal} {\bibinfo  {journal} {JHEP}\ }\textbf {\bibinfo {volume} {07}},\ \bibinfo {pages} {056}},\ \Eprint {https://arxiv.org/abs/1204.6184} {arXiv:1204.6184 [hep-lat]} \BibitemShut {NoStop}%
\bibitem [{\citenamefont {Hart}\ \emph {et~al.}(2000)\citenamefont {Hart}, \citenamefont {Laine},\ and\ \citenamefont {Philipsen}}]{Hart:2000ha}%
  \BibitemOpen
  \bibfield  {author} {\bibinfo {author} {\bibfnamefont {A.}~\bibnamefont {Hart}}, \bibinfo {author} {\bibfnamefont {M.}~\bibnamefont {Laine}},\ and\ \bibinfo {author} {\bibfnamefont {O.}~\bibnamefont {Philipsen}},\ }\bibfield  {title} {\bibinfo {title} {{Static correlation lengths in QCD at high temperatures and finite densities}},\ }\href {https://doi.org/10.1016/S0550-3213(00)00418-1} {\bibfield  {journal} {\bibinfo  {journal} {Nucl. Phys. B}\ }\textbf {\bibinfo {volume} {586}},\ \bibinfo {pages} {443} (\bibinfo {year} {2000})},\ \Eprint {https://arxiv.org/abs/hep-ph/0004060} {arXiv:hep-ph/0004060} \BibitemShut {NoStop}%
\bibitem [{\citenamefont {Bak}\ \emph {et~al.}(2007)\citenamefont {Bak}, \citenamefont {Karch},\ and\ \citenamefont {Yaffe}}]{Bak:2007fk}%
  \BibitemOpen
  \bibfield  {author} {\bibinfo {author} {\bibfnamefont {D.}~\bibnamefont {Bak}}, \bibinfo {author} {\bibfnamefont {A.}~\bibnamefont {Karch}},\ and\ \bibinfo {author} {\bibfnamefont {L.~G.}\ \bibnamefont {Yaffe}},\ }\bibfield  {title} {\bibinfo {title} {{Debye screening in strongly coupled N=4 supersymmetric Yang-Mills plasma}},\ }\href {https://doi.org/10.1088/1126-6708/2007/08/049} {\bibfield  {journal} {\bibinfo  {journal} {JHEP}\ }\textbf {\bibinfo {volume} {08}},\ \bibinfo {pages} {049}},\ \Eprint {https://arxiv.org/abs/0705.0994} {arXiv:0705.0994 [hep-th]} \BibitemShut {NoStop}%
\bibitem [{\citenamefont {Maezawa}\ \emph {et~al.}(2010)\citenamefont {Maezawa}, \citenamefont {Aoki}, \citenamefont {Ejiri}, \citenamefont {Hatsuda}, \citenamefont {Ishii}, \citenamefont {Kanaya}, \citenamefont {Ukita},\ and\ \citenamefont {Umeda}}]{Maezawa:2010vj}%
  \BibitemOpen
  \bibfield  {author} {\bibinfo {author} {\bibfnamefont {Y.}~\bibnamefont {Maezawa}}, \bibinfo {author} {\bibfnamefont {S.}~\bibnamefont {Aoki}}, \bibinfo {author} {\bibfnamefont {S.}~\bibnamefont {Ejiri}}, \bibinfo {author} {\bibfnamefont {T.}~\bibnamefont {Hatsuda}}, \bibinfo {author} {\bibfnamefont {N.}~\bibnamefont {Ishii}}, \bibinfo {author} {\bibfnamefont {K.}~\bibnamefont {Kanaya}}, \bibinfo {author} {\bibfnamefont {N.}~\bibnamefont {Ukita}},\ and\ \bibinfo {author} {\bibfnamefont {T.}~\bibnamefont {Umeda}} (\bibinfo {collaboration} {WHOT-QCD}),\ }\bibfield  {title} {\bibinfo {title} {{Electric and Magnetic Screening Masses at Finite Temperature from Generalized Polyakov-Line Correlations in Two-flavor Lattice QCD}},\ }\href {https://doi.org/10.1103/PhysRevD.81.091501} {\bibfield  {journal} {\bibinfo  {journal} {Phys. Rev. D}\ }\textbf {\bibinfo {volume} {81}},\ \bibinfo {pages} {091501} (\bibinfo {year} {2010})},\ \Eprint {https://arxiv.org/abs/1003.1361} {arXiv:1003.1361 [hep-lat]} \BibitemShut
  {NoStop}%
\bibitem [{\citenamefont {Plumari}\ \emph {et~al.}(2011)\citenamefont {Plumari}, \citenamefont {Alberico}, \citenamefont {Greco},\ and\ \citenamefont {Ratti}}]{Plumari:2011mk}%
  \BibitemOpen
  \bibfield  {author} {\bibinfo {author} {\bibfnamefont {S.}~\bibnamefont {Plumari}}, \bibinfo {author} {\bibfnamefont {W.~M.}\ \bibnamefont {Alberico}}, \bibinfo {author} {\bibfnamefont {V.}~\bibnamefont {Greco}},\ and\ \bibinfo {author} {\bibfnamefont {C.}~\bibnamefont {Ratti}},\ }\bibfield  {title} {\bibinfo {title} {{Recent thermodynamic results from lattice QCD analyzed within a quasi-particle model}},\ }\href {https://doi.org/10.1103/PhysRevD.84.094004} {\bibfield  {journal} {\bibinfo  {journal} {Phys. Rev. D}\ }\textbf {\bibinfo {volume} {84}},\ \bibinfo {pages} {094004} (\bibinfo {year} {2011})},\ \Eprint {https://arxiv.org/abs/1103.5611} {arXiv:1103.5611 [hep-ph]} \BibitemShut {NoStop}%
\bibitem [{\citenamefont {Kovtun}\ \emph {et~al.}(2005)\citenamefont {Kovtun}, \citenamefont {Son},\ and\ \citenamefont {Starinets}}]{Kovtun:2004de}%
  \BibitemOpen
  \bibfield  {author} {\bibinfo {author} {\bibfnamefont {P.}~\bibnamefont {Kovtun}}, \bibinfo {author} {\bibfnamefont {D.~T.}\ \bibnamefont {Son}},\ and\ \bibinfo {author} {\bibfnamefont {A.~O.}\ \bibnamefont {Starinets}},\ }\bibfield  {title} {\bibinfo {title} {{Viscosity in strongly interacting quantum field theories from black hole physics}},\ }\href {https://doi.org/10.1103/PhysRevLett.94.111601} {\bibfield  {journal} {\bibinfo  {journal} {Phys. Rev. Lett.}\ }\textbf {\bibinfo {volume} {94}},\ \bibinfo {pages} {111601} (\bibinfo {year} {2005})},\ \Eprint {https://arxiv.org/abs/hep-th/0405231} {arXiv:hep-th/0405231} \BibitemShut {NoStop}%
\bibitem [{\citenamefont {Altenkort}\ \emph {et~al.}(2023)\citenamefont {Altenkort}, \citenamefont {Eller}, \citenamefont {Francis}, \citenamefont {Kaczmarek}, \citenamefont {Mazur}, \citenamefont {Moore},\ and\ \citenamefont {Shu}}]{Altenkort:2022yhb}%
  \BibitemOpen
  \bibfield  {author} {\bibinfo {author} {\bibfnamefont {L.}~\bibnamefont {Altenkort}}, \bibinfo {author} {\bibfnamefont {A.~M.}\ \bibnamefont {Eller}}, \bibinfo {author} {\bibfnamefont {A.}~\bibnamefont {Francis}}, \bibinfo {author} {\bibfnamefont {O.}~\bibnamefont {Kaczmarek}}, \bibinfo {author} {\bibfnamefont {L.}~\bibnamefont {Mazur}}, \bibinfo {author} {\bibfnamefont {G.~D.}\ \bibnamefont {Moore}},\ and\ \bibinfo {author} {\bibfnamefont {H.-T.}\ \bibnamefont {Shu}},\ }\bibfield  {title} {\bibinfo {title} {{Viscosity of pure-glue QCD from the lattice}},\ }\href {https://doi.org/10.1103/PhysRevD.108.014503} {\bibfield  {journal} {\bibinfo  {journal} {Phys. Rev. D}\ }\textbf {\bibinfo {volume} {108}},\ \bibinfo {pages} {014503} (\bibinfo {year} {2023})},\ \Eprint {https://arxiv.org/abs/2211.08230} {arXiv:2211.08230 [hep-lat]} \BibitemShut {NoStop}%
\bibitem [{\citenamefont {Astrakhantsev}\ \emph {et~al.}(2017)\citenamefont {Astrakhantsev}, \citenamefont {Braguta},\ and\ \citenamefont {Kotov}}]{Astrakhantsev:2017nrs}%
  \BibitemOpen
  \bibfield  {author} {\bibinfo {author} {\bibfnamefont {N.}~\bibnamefont {Astrakhantsev}}, \bibinfo {author} {\bibfnamefont {V.}~\bibnamefont {Braguta}},\ and\ \bibinfo {author} {\bibfnamefont {A.}~\bibnamefont {Kotov}},\ }\bibfield  {title} {\bibinfo {title} {{Temperature dependence of shear viscosity of $SU(3)$--gluodynamics within lattice simulation}},\ }\href {https://doi.org/10.1007/JHEP04(2017)101} {\bibfield  {journal} {\bibinfo  {journal} {JHEP}\ }\textbf {\bibinfo {volume} {04}},\ \bibinfo {pages} {101}},\ \Eprint {https://arxiv.org/abs/1701.02266} {arXiv:1701.02266 [hep-lat]} \BibitemShut {NoStop}%
\bibitem [{\citenamefont {Meyer}(2007)}]{Meyer:2007ic}%
  \BibitemOpen
  \bibfield  {author} {\bibinfo {author} {\bibfnamefont {H.~B.}\ \bibnamefont {Meyer}},\ }\bibfield  {title} {\bibinfo {title} {{A Calculation of the shear viscosity in SU(3) gluodynamics}},\ }\href {https://doi.org/10.1103/PhysRevD.76.101701} {\bibfield  {journal} {\bibinfo  {journal} {Phys. Rev. D}\ }\textbf {\bibinfo {volume} {76}},\ \bibinfo {pages} {101701} (\bibinfo {year} {2007})},\ \Eprint {https://arxiv.org/abs/0704.1801} {arXiv:0704.1801 [hep-lat]} \BibitemShut {NoStop}%
\bibitem [{\citenamefont {Bors{\'a}nyi}\ \emph {et~al.}(2018)\citenamefont {Bors{\'a}nyi}, \citenamefont {Fodor}, \citenamefont {Giordano}, \citenamefont {Katz}, \citenamefont {Pasztor}, \citenamefont {Ratti}, \citenamefont {Sch{\"a}fer}, \citenamefont {Szabo},\ and\ \citenamefont {T{\'o}th}}]{Borsanyi:2018srz}%
  \BibitemOpen
  \bibfield  {author} {\bibinfo {author} {\bibfnamefont {S.}~\bibnamefont {Bors{\'a}nyi}}, \bibinfo {author} {\bibfnamefont {Z.}~\bibnamefont {Fodor}}, \bibinfo {author} {\bibfnamefont {M.}~\bibnamefont {Giordano}}, \bibinfo {author} {\bibfnamefont {S.~D.}\ \bibnamefont {Katz}}, \bibinfo {author} {\bibfnamefont {A.}~\bibnamefont {Pasztor}}, \bibinfo {author} {\bibfnamefont {C.}~\bibnamefont {Ratti}}, \bibinfo {author} {\bibfnamefont {A.}~\bibnamefont {Sch{\"a}fer}}, \bibinfo {author} {\bibfnamefont {K.~K.}\ \bibnamefont {Szabo}},\ and\ \bibinfo {author} {\bibfnamefont {B.}~\bibnamefont {T{\'o}th}},\ }\bibfield  {title} {\bibinfo {title} {{High statistics lattice study of stress tensor correlators in pure $SU(3)$ gauge theory}},\ }\href {https://doi.org/10.1103/PhysRevD.98.014512} {\bibfield  {journal} {\bibinfo  {journal} {Phys. Rev. D}\ }\textbf {\bibinfo {volume} {98}},\ \bibinfo {pages} {014512} (\bibinfo {year} {2018})},\ \Eprint {https://arxiv.org/abs/1802.07718} {arXiv:1802.07718 [hep-lat]} \BibitemShut
  {NoStop}%
\bibitem [{\citenamefont {Zhang}\ \emph {et~al.}(2025)\citenamefont {Zhang}, \citenamefont {Ding},\ and\ \citenamefont {Shu}}]{Zhang2025}%
  \BibitemOpen
  \bibfield  {author} {\bibinfo {author} {\bibfnamefont {C.}~\bibnamefont {Zhang}}, \bibinfo {author} {\bibfnamefont {H.-T.}\ \bibnamefont {Ding}},\ and\ \bibinfo {author} {\bibfnamefont {H.-T.}\ \bibnamefont {Shu}},\ }\href@noop {} {\bibinfo {title} {Shear and bulk viscosities of gluon plasma across the transition temperature from lattice qcd}},\ \bibinfo {howpublished} {Talk at Spicy Gluon 2025, Qingdao, Shandong} (\bibinfo {year} {2025})\BibitemShut {NoStop}%
\bibitem [{\citenamefont {Marty}\ \emph {et~al.}(2013)\citenamefont {Marty}, \citenamefont {Bratkovskaya}, \citenamefont {Cassing}, \citenamefont {Aichelin},\ and\ \citenamefont {Berrehrah}}]{Marty:2013ita}%
  \BibitemOpen
  \bibfield  {author} {\bibinfo {author} {\bibfnamefont {R.}~\bibnamefont {Marty}}, \bibinfo {author} {\bibfnamefont {E.}~\bibnamefont {Bratkovskaya}}, \bibinfo {author} {\bibfnamefont {W.}~\bibnamefont {Cassing}}, \bibinfo {author} {\bibfnamefont {J.}~\bibnamefont {Aichelin}},\ and\ \bibinfo {author} {\bibfnamefont {H.}~\bibnamefont {Berrehrah}},\ }\bibfield  {title} {\bibinfo {title} {{Transport coefficients from the Nambu-Jona-Lasinio model for $SU(3)_f$}},\ }\href {https://doi.org/10.1103/PhysRevC.88.045204} {\bibfield  {journal} {\bibinfo  {journal} {Phys. Rev. C}\ }\textbf {\bibinfo {volume} {88}},\ \bibinfo {pages} {045204} (\bibinfo {year} {2013})},\ \Eprint {https://arxiv.org/abs/1305.7180} {arXiv:1305.7180 [hep-ph]} \BibitemShut {NoStop}%
\bibitem [{\citenamefont {Ghiglieri}\ \emph {et~al.}(2018)\citenamefont {Ghiglieri}, \citenamefont {Moore},\ and\ \citenamefont {Teaney}}]{Ghiglieri:2018dib}%
  \BibitemOpen
  \bibfield  {author} {\bibinfo {author} {\bibfnamefont {J.}~\bibnamefont {Ghiglieri}}, \bibinfo {author} {\bibfnamefont {G.~D.}\ \bibnamefont {Moore}},\ and\ \bibinfo {author} {\bibfnamefont {D.}~\bibnamefont {Teaney}},\ }\bibfield  {title} {\bibinfo {title} {{QCD Shear Viscosity at (almost) NLO}},\ }\href {https://doi.org/10.1007/JHEP03(2018)179} {\bibfield  {journal} {\bibinfo  {journal} {JHEP}\ }\textbf {\bibinfo {volume} {03}},\ \bibinfo {pages} {179}},\ \Eprint {https://arxiv.org/abs/1802.09535} {arXiv:1802.09535 [hep-ph]} \BibitemShut {NoStop}%
\end{thebibliography}%
\end{document}